\documentclass[aps,prc,showpacs,showkeys,nofootinbib]{revtex4}

\usepackage{graphicx}
\usepackage{dcolumn}
\usepackage{amsthm}
\usepackage{bm}

\newcommand{\rhobf}{\mbox{\boldmath $\rho$}}
\newcommand{\rhobfsm}{\small \mbox{\boldmath $\rho$}}

\begin{document}

\title
{Study of compound nucleus formation via bremsstrahlung emission in proton $\alpha$-particle scattering}

\author{Sergei~P.~Maydanyuk}%
\email{maidan@kinr.kiev.ua}%
\affiliation{Institute of Modern Physics, Chinese Academy of Sciences, Lanzhou, 730000, China}
\affiliation{Institute for Nuclear Research, National Academy of Sciences of Ukraine, Kiev, 03680, Ukraine}
\author{Peng-Ming~Zhang}%
\email{zhpm@impcas.ac.cn}%
\affiliation{Institute of Modern Physics, Chinese Academy of Sciences, Lanzhou, 730000, China}
\affiliation{State Key Laboratory of Theoretical Physics, Institute of Theoretical Physics, Chinese Academy of Sciences, Beijing 100190, China}











\date{\small\today}

\begin{abstract}
In this paper a role of many-nucleon dynamics in formation of the compound $^{5}{\rm Li}$ nucleus in the scattering of protons off $\alpha$-particles at the proton incident energies up to 20~MeV is investigated.
We propose a bremsstrahlung model allowing to extract information about probabilities of formation of such nucleus on the basis of analysis of experimental cross-sections of the bremsstrahlung photons.
In order to realize this approach, the model includes elements of microscopic theory and also probabilities of formation of the short-lived compound nucleus.
Results of calculations of the bremsstrahlung spectra are in good agreement with the experimental cross-sections.
%
\end{abstract}


\pacs{%
41.60.-m, 
25.40.Cm, 
03.65.Xp, 
25.70.Gh, 
24.10.Ht, 
23.20.Js} 


\keywords{
bremsstrahlung,
photon,
scattering of protons off alpha particles,
microscopic model,
tunneling,
compound nucleus,
fusion,
$^{5}{\rm Li}$}

\maketitle

\section{Introduction
\label{sec.inroduction}}

Understanding of nuclear interactions between nucleons is hot topic of physics.
In such a line, bremsstrahlung photons emitted during nuclear reactions provide important information about dynamics of nucleons of composed nuclear system,
which is formed on the basis of interactions.
The bremsstrahlung of the emitted photons in nuclear reactions where nuclei are described on the microscopic level,
has been studied for a long time%
~\cite{Liu.1981.PRC.v23,Baye.1985.NPA,Liu.1990.PRC.v41,Liu.1990.PRC.v42,
Baye.1991.NPA,Liu.1992.FBS,Dohet-Eraly.2011.JPCS,Dohet-Eraly.2011.PRC,Dohet-Eraly.2013.PhD,
Dohet-Eraly.2013.JPCS,Dohet-Eraly.2013.PRC,Dohet-Eraly.2014.PRC.v89,Dohet-Eraly.2014.PRC.v90}.
In such frameworks, bremsstrahlung photons are successfully applied for deeper understanding of nucleon-nucleon interactions.
The nucleon-nucleon scattering data (for example, Nijmegen data set \cite{Stoks.1993.PRC}) and properties of the deuteron are used for controlling parameters of the microscopical cluster models.

A crucial point in description of scattering properties is accurate (precise) determination of wave functions for scattering states of the nuclear system.
In their determination, scattering theory and theory of nuclear reactions have traditional way, where nuclei from lightest up to super-heavy are included into consideration.
In such a frameworks, interactions between the scattering objects are described often in potential approach, which parameters are extracted from existed experimental data.
Classical objects for study here are effects of collective motion, strong quantum phenomena, processes of fusion and breakup inside the nuclear system, etc.

Results of such a work are used as a basis for construction of other bremsstrahlung models in nuclear physics.
Our bremsstrahlung model is developed along such a line
\cite{Maydanyuk.2003.PTP,Maydanyuk.2006.EPJA,Maydanyuk.2008.EPJA,Giardina.2008.MPLA,Maydanyuk.2009.NPA,
Maydanyuk.2009.JPS,Maydanyuk.2009.TONPPJ,Maydanyuk.2010.PRC,Maydanyuk.2011.JPG,Maydanyuk.2011.JPCS,Maydanyuk.2012.PRC,Maydanyuk_Zhang.2015.PRC,Maydanyuk_Zhang_Zou.2016.PRC}.
According to its logic, determination of the wave functions for the scattering states is quantum mechanical task with interacting potentials, that allows to keep maximally accurate quantum effect
(that can be interesting in cases where they are strong, for example, see~\cite{Maydanyuk_Zhang.2015.NPA}).
As attractive point of such a line,
inverse scattering theory~\cite{Zakhariev.1985.book} can be naturally added, to achieve exact coincidence between experimental and calculated cross-sections (that provides information about the corresponding potentials).
Our approach was successfully tested on set of tasks where experimental data on the emission of the bremsstrahlung photons in nuclear reactions is essentially larger.
These are experimental data for
$\alpha$-decay~\cite{Boie.2007.PRL,Boie.2009.PhD,Maydanyuk.2008.EPJA,Giardina.2008.MPLA,D'Arrigo.1994.PHLTA,Kasagi.1997.JPHGB,Kasagi.1997.PRLTA}
(see also some calculations \cite{Batkin.1986.SJNCA,Dyakonov.1996.PRLTA,Papenbrock.1998.PRLTA,Tkalya.1999.JETP,
Tkalya.1999.PHRVA,Bertulani.1999.PHRVA,Takigawa.1999.PHRVA,Flambaum.1999.PRLTA,Dyakonov.1999.PHRVA,So_Kim.2000.JKPS,
Misicu.2001.JPHGB,Dijk.2003.FBSSE,Ohtsuki.2006.CzJP,Amusia.2007.JETP,Jentschura.2008.PRC}),
scattering of protons off nuclei~\cite{Edington.1966.NP,Koehler.1967.PRL,Kwato_Njock.1988.PLB,Pinston.1989.PLB,
Pinston.1990.PLB,Clayton.1991.PhD,Clayton.1992.PRC.p1810,Clayton.1992.PRC.p1815,Chakrabarty.1999.PRC,Goethem.2002.PRL}
(see also reviews on situation in such a research \cite{Kamanin.1989.PEPAN,Pluiko.1987.PEPAN}),
spontaneous fission~\cite{Ploeg.1995.PRC,Kasagi.1989.JPSJ,Luke.1991.PRC,Hofman.1993.PRC,Varlachev.2007.BRASP,Eremin.2010.IJMPE,Pandit.2010.PLB},
etc.
(see also predicted spectra for emission of protons from nuclei~\cite{Kurgalin.2001.IRAN,Maydanyuk.2011.JPG,Maydanyuk.2012.PRC,Maydanyuk_Zhang.2015.PRC},
ternary fission~\cite{Maydanyuk.2011.JPCS}).


So, all existed variety of experimental data of emission of the bremsstrahlung photons in nuclear reactions can be separated into two following groups.
To the first group one can include the data~\cite{Wolfli.1971.PRL,Peyer.1971.PLB,Frois.1973.PRC,Anzelon.1975.NPA}, analyzed in the microscopic cluster models developed formalism of
resonating group method~\cite{Wildermuth.1977.book,Tang.1978.PR,Tang.1981.lectures}
and generator coordinate method~\cite{Horiuchi.1977.PTPS}.
Such a data are obtained in the coplanar geometry, where maxima in the bremsstrahlung cross-sections are observed at some energies (i.e. resonances).
Usually, such maxima are explained by resonant states of the compound nuclear system.
In frameworks of such a formalism, kinematic relations between energy of the emitted photon and energy of relative motion of two nuclei or nucleus and nucleon are applied.


The experimental data of the second group are obtained relatively energy of the emitted photons
(here, in calculations usually kinematic relations between energy of the emitted photons and energy of relative motion between two nuclei or nucleus and nucleon are not imposed).
Such cross-sections are obtained in the large regions of energy, they have smooth continuous shapes where any sharp resonant peaks are not observed.
However, potentials of interactions usually do not include explicitly such resonant states of the compound nuclear system at definite energies.
%
Absence of the fixed dependence between the energy of the emitted photons and the energy of the relative motion between two nuclei or nucleus and nucleon is
justified by that such experimental bremsstrahlung cross-sections for different types of nuclear decays and the scattering of protons off nuclei at fixed incident proton energy are continuous, usually have no resonant peaks and are presented inside enough large region of the emitted photon energy.
In description of such an experimental data our approach proves oneself enough accurate (successfully).
%
From such a point of view, it could be interesting for us to investigate the experimental data of the first group on the basis of our formalism.
We see that understanding of the nuclear interactions on the basis of many nucleon forces, is one of the most important tasks in nuclear physics.
Therefore, in order to base possibility to study the interactions between nucleons in our formalism, we include some elements of the microscopic theory into it.
These ideas represent a main aim of this paper.

%
%

Another attractive advance of our approach is possibility to include into operator of emission the additional terms, connected spinor properties of nucleons with their momenta (i.e. possibility to study influence of dynamics and spins of nucleons (their interference) on the emission of photons).
Such corrections are appeared in result of many-nucleon generalization of Dirac equation and the corresponding hamiltonian of the nuclear system
(where the first approximation is many-nucleon Pauli equation, see~\cite{Maydanyuk.2012.PRC} for details).
So, thy have another origin than appearance of the spin states in the resonating group method and the generator coordinate method.
In particular, in frameworks of such a model \cite{Maydanyuk_Zhang.2015.PRC}, we explain the hump-shaped plateau in the intermediate and high energy regions of
the experimental bremsstrahlung data~\cite{Goethem.2002.PRL} by essential presence of the incoherent emission
(formed by interactions between the scattering proton and nucleus with the internal many-nucleon structure)
for the scattering of $p + ^{208}{\rm Pb}$ at the proton incident energies of 140 and 145~MeV,
and the scattering of $p + ^{12}{\rm C}$, $p + ^{58}{\rm Ni}$, $p + ^{107}{\rm Ag}$ and $p + ^{197}{\rm Au}$ at the proton incident energy of 190~MeV,
while at low energies the coherent emission
(formed by interaction between the scattering proton and nucleus as a whole without internal many-nucleon structure)
predominates which produces the logarithmic shape spectrum.
Such an approach allows to develop theory for description of the bremsstrahlung photons emitted in the nuclear reactions in the relativistic region of energies.

Else one useful advance of our model is description of quantum effects shown during fusion and break-up processes inside the compound nuclear system.
According to~\cite{Maydanyuk_Zhang.2015.NPA}, quantum effects participating in capture of the $\alpha$ particle by the $^{40}{\rm Ca}$ and $^{44}{\rm Ca}$ nuclei, are not small, and inclusion of their description into the model allows to essentially improve agreement between the calculated spectra and experimental data
(this method found new parametrization of the $\alpha$--nucleus potential and fusion probabilities and decreased the error
by $41.72$ times for $\alpha + ^{40}{\rm Ca}$ and $34.06$ times for $\alpha + ^{44}{\rm Ca}$ in a description of experimental data~\cite{Eberhard.1979.PRL}
in comparison with previously existing results).
%
From this point of view, it could be interesting to include ideas of such a formalism of the processes of fusion and breakup into the bremsstrahlung model. But, in order to include formalism of many-nucleon interactions, we draw attention on one of light systems, which were studied previously and for which the experimental data exists.
This is the scattering of protons off the $\alpha$-particles,
for which the bremsstrahlung cross-sections were measured in the coplanar geometry~\cite{Wolfli.1971.PRL}.
Here, suitable object for investigations of processes of nuclear formation and breakup is the system from five nucleons
in form of the short-lived $^{5}{\rm Li}$ nucleus.
Analysis of such a task is also included into our research in this paper.

%
%

In Sec.~\ref{sec.2}, our improved bremsstrahlung model applied for the scattering of protons off alpha-particles is presented.
Here, after formulating of operator of emission of bremsstrahlung photons for such a nuclear system,
emphasis is made on construction of formalism for the matrix elements in the microscopic approach,
with next application of the multipolar expansion for wave function of photons
and obtaining the bremsstrahlung cross-sections.
In Sec.~\ref{sec.results} we perform theoretical analysis of the bremsstrahlung emission in this reaction.
Here,
after analysis of phase shifts of the scattering wave functions obtained in our approach
and comparing them with empirical data~\cite{Arndt.1971.PRC},
we calculate contributions of the emitted photons from different transitions between states,
obtain the full bremsstrahlung spectrum comparing it with experimental data~\cite{Wolfli.1971.PRL} of Arndt, Roper, and Shotwell, and
calculations of Liu, Tang and Kanada~\cite{Liu.1990.PRC.v42},
and Dohet-Eraly~\cite{Dohet-Eraly.2014.PRC.v89}.
Concluding remarks are presented in Sec.~\ref{sec.conclusions}.
Calculations of one-nucleon space matrix elements over bound states of nucleus are added in Appendixes~\ref{sec.app.1} and \ref{sec.app.2},
for convenience.


\section{Model
\label{sec.2}}

\subsection{Operator of emission and wave function of the proton - nucleus system
\label{sec.2.1}}

We shall start from the leading form Eq.~(7) of the photon emission operator $\hat{H}_{\gamma}$ in \cite{Maydanyuk_Zhang.2015.PRC,Maydanyuk_Zhang_Zou.2016.PRC},
generalizing it for the system of the scattering proton and nucleus composed of $A$ nucleons
in the laboratory system.
Using presentation for the vector potential of the electromagnetic field in form (5) in \cite{Maydanyuk.2012.PRC},
we obtain
\begin{equation}
\begin{array}{lcl}
  \hat{H}_{\gamma} =
    -\,e\, \sqrt{\displaystyle\frac{2\pi}{w_{\rm ph}}}\,
    \displaystyle\sum\limits_{\alpha=1,2} \mathbf{e}^{(\alpha),*}\;
    \biggl\{
      \displaystyle\frac{z_{\rm p}}{m_{\rm p}}\; e^{-i \mathbf{kr}_{\rm p}}\, \mathbf{p}_{\rm p} +
      \displaystyle\sum\limits_{j=1}^{A}
        \displaystyle\frac{z_{j}}{m_{j}}\; e^{-i \mathbf{kr}_{j}}\, \mathbf{p}_{j}
    \Bigr\}.
\end{array}
\label{eq.2.1.1}
\end{equation}
%
%
%
%
%
%
Here,
star denotes complex conjugation,
$z_{j}$ and $m_{j}$ are the electromagnetic charge and mass of the nucleon with number $j$,
$m_{\rm p}$ is mass of proton,
$\mathbf{p}_{j} = -i\hbar\, \mathbf{d}/\mathbf{dr}_{j} $ is the momentum operator for the nucleon with number $j$
(we number nucleons of the nucleus by index $j$).
$\mathbf{e}^{(\alpha)}$ are unit vectors of the polarization of the photon emitted [$\mathbf{e}^{(\alpha), *} = \mathbf{e}^{(\alpha)}$], $\mathbf{k}$ is the wave vector of the photon and $w_{\rm ph} = k c = \bigl| \mathbf{k} \bigr|\: c$. Vectors $\mathbf{e}^{(\alpha)}$ are perpendicular to $\mathbf{k}$ in the Coulomb gauge. We have two independent polarizations $\mathbf{e}^{(1)}$ and $\mathbf{e}^{(2)}$ for the photon with momentum $\mathbf{k}$ ($\alpha=1,2$).
In this paper we shall use the system of units where $\hbar = 1$ and $c = 1$.

Now we turn to the center-of-mass frame.
We define coordinate of centers of masses
for the nucleus as $\mathbf{R}_{A}$ and
for the complete system as $\mathbf{R}$ having form
$\mathbf{R}_{A} = \sum_{j=1}^{A} m_{j}\, \mathbf{r}_{A j} / m_{A}$,
$\mathbf{R} = (m_{A}\mathbf{R}_{A} + m_{\rm p}\mathbf{r}_{\rm p}) / (m_{A}+m_{\rm p})$,
where $m_{\rm p}$ and $m_{A}$ are masses of the scattering proton and nucleus.
Introducing new relative coordinates $\rhobf_{A j}$ and $\mathbf{r}$ as
$\mathbf{r}_{j} = \mathbf{R}_{A} + \rhobf_{A j}$,
$\mathbf{r} = \mathbf{r}_{\rm p} - \mathbf{R}_{A}$,
we find the corresponding momenta
$\mathbf{p}_{j} = \mathbf{P}_{A} + \mathbf{\tilde{p}}_{A j}$,
$\mathbf{p} = \mathbf{p}_{\rm p} - \mathbf{P}_{A}$,
where
$\mathbf{p}_{\rm p} = -\,i\hbar\, \mathbf{d} / \mathbf{dr}_{\rm p}$,
$\mathbf{P}_{A} = -\,i\hbar\, \mathbf{d} / \mathbf{dR}_{A}$,
$\mathbf{\tilde{p}}_{A j} = -i\hbar\, \mathbf{d} / \mathbf{d\rho}_{A j}$.
Using these formulas, we obtain
\begin{equation}
\begin{array}{cccc}
  \mathbf{R}_{A} = \mathbf{R} - c_{\rm p}\, \mathbf{r}, &
  \mathbf{r}_{\rm p} = \mathbf{R} + c_{A}\, \mathbf{r}, %
  \mathbf{r}_{j} = \mathbf{R} - c_{\rm p}\, \mathbf{r} + \rhobf_{A j},
\end{array}
\label{eq.2.1.2}
\end{equation}
where we introduced $c_{A} = \frac{m_{A}}{m_{A}+m_{\rm p}}$ and $c_{\rm p} = \frac{m_{\rm p}}{m_{A}+m_{\rm p}}$.
Substituting these expressions to eq.~(\ref{eq.2.1.1}), we find%
%
\begin{equation}
\begin{array}{lcl}
  \vspace{0mm}
  \hat{H}_{\gamma} \;\; = \;\;
  -\,e\; \sqrt{\displaystyle\frac{2\pi}{w_{\rm ph}}}\,
    \displaystyle\sum\limits_{\alpha=1,2} \mathbf{e}^{(\alpha),*}\;
    e^{-i \mathbf{k} \bigl[\mathbf{R} - c_{\rm p} \mathbf{r} \bigr]}\; 
    \Biggl\{
      \Bigl[
        e^{-i \mathbf{k}\mathbf{r}}\,
          \displaystyle\frac{z_{\rm p}}{m_{\rm p}}\, +\,

          \displaystyle\sum\limits_{j=1}^{A}  \displaystyle\frac{z_{j}}{m_{j}}\;
          e^{-i \mathbf{k} \rhobfsm_{A j} }
      \Bigr]\: \mathbf{P}\; + \\

  \;\; +\:
      \Bigl[
        c_{A}\, e^{-i \mathbf{k}\mathbf{r}}
          \displaystyle\frac{z_{\rm p}}{m_{\rm p}}\, -
        c_{\rm p}
          \displaystyle\sum\limits_{j=1}^{A} \displaystyle\frac{z_{j}}{m_{j}}\;
          e^{-i \mathbf{k} \rhobf_{A j} }
      \Bigr]\: \mathbf{p}\; + 

      \displaystyle\sum\limits_{j=1}^{A}
        \displaystyle\frac{z_{j}}{m_{j}}\;
        e^{-i \mathbf{k} \rhobf_{A j}}\, \mathbf{\tilde{p}}_{A j}
    \Biggr\}.
\end{array}
\label{eq.2.1.3}
\end{equation}

We define the wave function of the full nuclear system as
\begin{equation}
  \Psi (1, 2 \ldots A+1) =
  \mathcal{A}\, \bigl[ \psi_{\lambda_{1}} (1), \psi_{\lambda_{2}} (2) \ldots \psi_{\lambda_{A+1}} (A+1) \bigr],
\label{eq.2.3.1}
\end{equation}
with
$\mathcal{A}$ being an antisymmetrization operator.
One-nucleon functions $\psi_{\lambda_{s}}(s)$ represent the multiplication of space and spin-isospin functions as
\begin{equation}
  \psi_{\lambda_{s}} (s) =
  \varphi_{n_{s}} (\mathbf{r}_{s})\,
  \bigl|\, \sigma^{(s)} \tau^{(s)} \bigr\rangle,
\label{eq.2.3.2}
\end{equation}
where
$\varphi_{n_{s}}$ is space function of the nucleon with number $s$,
$n_{s}$ is number of state of the space function of the nucleon with number $s$,
$\bigl|\, \sigma^{(s)} \tau^{(s)} \bigr\rangle$ is spin-isospin function of the nucleon with number $s$.

\subsection{Matrix element of emission
\label{sec.2.4}}

We shall assume
$\Phi_{\bar{s}} (\mathbf{R}) =  e^{-i\,\mathbf{K}_{\bar{s}}\cdot\mathbf{R}}$
%
%
%
where $\bar{s} = i$ or $f$ (indexes $i$ and $f$ denote the initial state, i.e. the state before emission of photon,
and the final state, i.e. the state after emission of photon),
$\mathbf{K}_{s}$ is momentum of the total system~\cite{Kopitin.1997.YF}.
Suggesting
%
$
  \mathbf{K}_{i} = 0,
$
%
we calculate the matrix element:
\begin{equation}
\begin{array}{lcl}
  \vspace{0mm}
  \langle \Psi_{f} |\, \hat{H}_{\gamma} |\, \Psi_{i} \rangle  \;\; = \;\;
  -\,\displaystyle\frac{e}{m_{\rm p}}\; \sqrt{\displaystyle\frac{2\pi}{w_{\rm ph}}}\,
    \displaystyle\sum\limits_{\alpha=1,2} \mathbf{e}^{(\alpha),*}\;
  \Bigl\{ M_{1} + M_{2} + M_{3} \Bigr\},
\end{array}
\label{eq.2.4.1}
\end{equation}
where
\begin{equation}
\begin{array}{lcl}
  M_{1} =
  \biggl\langle
    \Psi_{f}\,
  \biggl|\,
    e^{i\,(\mathbf{K}_{f} - \mathbf{k})\cdot\mathbf{R}}\:
    e^{i\, c_{\alpha} \mathbf{kr} }\; 

      \Bigl[
        e^{-i \mathbf{k}\mathbf{r}}\, z_{\rm p}\, +
          \displaystyle\sum\limits_{j=1}^{A} z_{j}\, \displaystyle\frac{m_{\rm p}}{m_{j}}\;
          e^{-i \mathbf{k} \rhobf_{A j} }
      \Bigr]\: \mathbf{P}\;
    \biggr|\,
      \Psi_{i}
    \biggr\rangle, \\

  M_{2} =
    \biggl\langle
      \Psi_{f}\,
    \biggl|\,
      e^{i\,(\mathbf{K}_{f} - \mathbf{k})\cdot\mathbf{R}}\:
      e^{i\, c_{\alpha} \mathbf{kr} }\,
      \Bigl[
          e^{-i \mathbf{k}\mathbf{r}}\, c_{A}\, z_{\rm p}\, - 

        c_{\alpha}\,
          \displaystyle\sum\limits_{j=1}^{A} z_{j}\,\displaystyle\frac{m_{\rm p}}{m_{j}}\;
          e^{-i \mathbf{k} \rhobf_{A j} }
      \Bigr]\: \mathbf{p}
    \biggr|\,
      \Psi_{i}\,
    \biggr\rangle, \\

  M_{3} =
    \biggl\langle
      \Psi_{f}\,
    \biggl|\,
      e^{i\,(\mathbf{K}_{f} - \mathbf{k})\cdot\mathbf{R}}\:
      e^{i\, c_{\alpha} \mathbf{kr} }\,
      \displaystyle\sum\limits_{j=1}^{A}
        z_{j}\,\displaystyle\frac{m_{\rm p}}{m_{j}}\;
        e^{-i \mathbf{k} \rhobf_{A j}}\, \mathbf{\tilde{p}}_{A j}\;
    \biggr|\,
      \Psi_{i}
    \biggr\rangle.
\end{array}
\label{eq.2.4.2}
\end{equation}
%
We will not use the first term $M_{1}$ (as we shall study decay in the center-of-mass system and neglect by possible response), and the third term $M_{3}$ (as we shall not study contribution of photon emission caused by the deformation of the daughter nucleus during decay).
So, we shall calculate the second matrix element.
Substituting representation for the wave function, we obtain:
\begin{equation}
\begin{array}{lcl}
  M_{2} =
    \delta(\mathbf{K}_{f} - \mathbf{k})\;
    \biggl\{
      c_{A}\,
    \Bigl\langle
      \psi_{f} (1 \cdots A)\,
    \Bigl|\,
      e^{i\, c_{\rm p} \mathbf{kr} }\,
        e^{-i \mathbf{k}\mathbf{r}}
        \: \mathbf{p}
    \Bigr|\,
      \psi_{i} (1 \cdots A)\,
    \Bigr\rangle\; - \\

  -\; c_{\rm p}\,
    \displaystyle\sum\limits_{j=1}^{Z_{A}}\,
    \Bigl\langle
      \psi_{f} (1 \cdots A)\,
    \Bigl|\,
      e^{i\, c_{\rm p} \mathbf{kr} }\,
      f_{Aj}\, (\rhobf_{A j})
      \: \mathbf{p}
    \Bigr|\,
      \psi_{i} (1 \cdots A)\,
    \Bigr\rangle
    \biggr\},
\end{array}
\label{eq.2.4.3}
\end{equation}
where
\begin{equation}
\begin{array}{lcl}
  f_{Aj}\, (\rhobf_{Aj}) =
    z_{j}\,\displaystyle\frac{m_{\rm p}}{m_{j}}\; e^{-i \mathbf{k} \rhobf_{A j}}.
\end{array}
\label{eq.2.4.4}
\end{equation}
Matrix element from operator, dependent on two nucleons with numbers $i$ and $j$, can be written in form of linear combination of two-nucleon matrix elements
as
%
%
\begin{equation}
\begin{array}{lcl}
\vspace{1mm}
  \langle \psi_{f} (1 \cdots A )\, |\, \hat{V} (\mathbf{r}_{i}, \mathbf{r}_{j})\,
    |\, \psi_{i} (1 \cdots A ) \rangle =  \\
 \vspace{1mm}
   = \quad
    \displaystyle\frac{1}{A\,(A-1)}\;
    \displaystyle\sum\limits_{k=1}^{A}
    \displaystyle\sum\limits_{m=1, m \ne k}^{A}
    \biggl\{
      \langle \psi_{k}(i)\, \psi_{m}(j) |\,
      \hat{V} (\mathbf{r}_{i}, \mathbf{r}_{j})\, |\, \psi_{k}(i)\, \psi_{m}(j) \rangle - 

    \langle \psi_{k}(i)\, \psi_{m}(j) |\,
    \hat{V} (\mathbf{r}_{i}, \mathbf{r}_{j})\, |\, \psi_{m}(i)\, \psi_{k}(j) \rangle
  \biggr\}.
\end{array}
\label{eq.2.4.5}
\end{equation}
Here, summation is performed over all states of nucleons.
But, if operator is not dependent on relative distances between nucleons and it does not act on spin and isospin states of nucleons, then it needs to calculate all terms of matrix element separately.
In result, we obtain:
%
%
\begin{equation}
  M_{2}\; =\; \delta(\mathbf{K}_{f} - \mathbf{k})\, \Bigl\{ M_{21} - M_{22} \Bigr\},
\label{eq.2.4.6}
\end{equation}
where
\begin{equation}
\begin{array}{lcl}
  M_{21} & = &
    \displaystyle\frac{c_{A}}{A+1}\;
    \displaystyle\sum\limits_{k=1}^{A+1}
    \Bigr\langle \psi_{k}(i_{\rm p})
    \Bigl|\,
      e^{i\, c_{\rm p} \mathbf{kr} }\,
        e^{-i \mathbf{k}\mathbf{r}}\,
        f_{\rm p} (\rhobf_{\rm p})
        \: \mathbf{p}
    \Bigr|\,
      \psi_{k}(i_{\rm p})
    \Bigr\rangle, \\

  M_{22} & = &
  c_{\rm p}\,
    \displaystyle\sum\limits_{j=1}^{Z_{A}}\,
    \displaystyle\frac{1}{A\,(A+1)}\;
    \displaystyle\sum\limits_{k=1}^{A+1}
    \displaystyle\sum\limits_{m=1, m \ne k}^{A+1}
    \biggl\{
    \Bigr\langle \psi_{k}(i_{\rm p}) \Bigl|\,
      e^{i\, c_{\rm p} \mathbf{kr} }\: \mathbf{p}
    \Bigr|\, \psi_{k}(i_{\rm p}) \Bigr\rangle\;
    \Bigr\langle \psi_{m}(j_{A}) \Bigl|\,
      f_{j}\, (\rhobf_{A j})
    \Bigr|\, \psi_{m}(j_{A}) \Bigr\rangle\; - \\

   & - &
    \Bigr\langle \psi_{k}(i_{\rm p}) \Bigl|\,
      e^{i\, c_{\rm p} \mathbf{kr} }\: \mathbf{p}
    \Bigr|\, \psi_{m}(i_{\rm p}) \Bigr\rangle\;
    \Bigr\langle \psi_{m}(j_{A}) \Bigl|\,
      f_{j}\, (\rhobf_{A j})
    \Bigr|\, \psi_{k}(j_{A}) \Bigr\rangle\;
  \biggr\}.
\end{array}
\label{eq.2.4.7}
\end{equation}

\subsection{Matrix element in multipole expansion of wave function of photons
\label{sec.2.7}}

After summation over spin-isospin states (see Appendix~\ref{sec.app.1}, for details),
in a case of the $\alpha$-particle as the target nucleus, we obtain:
\begin{equation}
\begin{array}{lcl}
\vspace{3mm}
  M_{21} & = &
  \displaystyle\frac{c_{A}}{5}\;
  \Bigl\{
    2\; i\, (c_{\rm p}-1)\, \mathbf{k}\;
      e^{-\, (c_{\rm p}-1)^{2}\, (a^{2} k_{x}^{2} + b^{2} k_{y}^{2} + c^{2} k_{z}^{2})\,/4}\; + \;
    J_{1} (\mathbf{k})
  \Bigr\}, \\

\vspace{1mm}
  M_{22} & = &
  \displaystyle\frac{c_{\rm p}}{10}\;
  \Bigl\{
    2\;J_{2} (\mathbf{k})\; e^{-\, (a^{2} k_{x}^{2} + b^{2} k_{y}^{2} + c^{2} k_{z}^{2})\,/4}\; +\;
    i\, c_{\rm p}\, \mathbf{k}\;
      J_{3} (\mathbf{k})\;
      e^{-\, c_{\rm p}^{2}\, (a^{2} k_{x}^{2} + b^{2} k_{y}^{2} + c^{2} k_{z}^{2})\,/4}
  \Bigr\},
\end{array}
\label{eq.2.7.1}
\end{equation}
where we introduced the integrals
\begin{equation}
\begin{array}{lcl}
\vspace{3mm}
  J_{1} (\mathbf{k})
  & = &
    \Bigr\langle \varphi_{k}(\rhobf) \Bigl|\,
      e^{i\, (c_{\rm p}-1) \mathbf{k}\rhobfsm }\,
      \mathbf{p}
    \Bigr|\, \varphi_{k}(\rhobf) \Bigr\rangle_{k=2 (scat. state)}, \\

\vspace{3mm}
  J_{2} (\mathbf{k})
  & = &
    \Bigr\langle \varphi_{k}(\rhobf) \Bigl|\,
      e^{i\, c_{\rm p} \mathbf{k}\rhobfsm }\,
      \mathbf{p}
    \Bigr|\, \varphi_{k}(\rhobf) \Bigr\rangle_{k=2 (scat. state)}
  \Bigr\}, \\

  J_{3} (\mathbf{k})
  & = &
    \Bigr\langle \varphi_{k}(\rhobf) \Bigl|\,
      e^{- i\, \mathbf{k}\rhobfsm }\,
    \Bigr|\, \varphi_{k}(\rhobf) \Bigr\rangle_{k=2 (scat. state)}
  \Bigr\}.
\end{array}
\label{eq.2.7.2}
\end{equation}
According to results of variational analysis for the $\alpha$-particle \cite{Steshenko.1971.YF}, it has spherically symmetric shape in the ground state
(we have $a = b = c =1.02$~fm) and eqs.~(\ref{eq.2.7.1}) are simplified as
\begin{equation}
\begin{array}{lcl}
\vspace{3mm}
  M_{21} & = &
  \displaystyle\frac{c_{A}}{5}\;
  \Bigl\{
    2\; i\, (c_{\rm p}-1)\, \mathbf{k}\;
      e^{-\, (c_{\rm p}-1)^{2}\, a^{2} k^{2}/4}\; + \;
    J_{1} (\mathbf{k})
  \Bigr\}, \\

\vspace{1mm}
  M_{22} & = &
  \displaystyle\frac{c_{\rm p}}{10}\;
  \Bigl\{
    2\;J_{2} (\mathbf{k})\; e^{- a^{2} k^{2}/4}\; +\;
    i\, c_{\rm p}\, \mathbf{k}\;
      J_{3} (\mathbf{k})\;
      e^{-\, c_{\rm p}^{2} a^{2} k^{2}/4}
  \Bigr\},
\end{array}
\label{eq.2.7.3}
\end{equation}
where $k^{2} = k_{x}^{2} + k_{y}^{2} + k_{z}^{2}$.
Now we calculate multiplications of these functions on vectors of polarization of photons. Taking into account that vectors $\mathbf{e}^{1,2}$ and $\mathbf{k}$ are perpendicular,
we have
property:
\begin{equation}
\begin{array}{lcl}
\vspace{3mm}
  \displaystyle\sum\limits_{\alpha=1,2} \mathbf{e}^{(\alpha),*}\;
  M_{21} & = &
  \displaystyle\frac{c_{A}}{5}\;
  \displaystyle\sum\limits_{\alpha=1,2} \mathbf{e}^{(\alpha),*}\;
    J_{1} (\mathbf{k}), \\

\vspace{1mm}
  \displaystyle\sum\limits_{\alpha=1,2} \mathbf{e}^{(\alpha),*}\;
  M_{22} & = &
  \displaystyle\frac{c_{\rm p}}{5}\;
  \displaystyle\sum\limits_{\alpha=1,2} \mathbf{e}^{(\alpha),*}\;
    J_{2} (\mathbf{k})\; e^{- a^{2} k^{2}/4}
\end{array}
\label{eq.2.7.4}
\end{equation}
and the full matrix element (\ref{eq.2.4.1}) is equal to
\begin{equation}
\begin{array}{lcl}
  \langle \Psi_{f} |\, \hat{H}_{\gamma} |\, \Psi_{i} \rangle & = &
  -\,\displaystyle\frac{e}{m_{\rm p}}\, \sqrt{\displaystyle\frac{2\pi}{w_{\rm ph}}} \cdot
    p_{fi}\;
    \delta(\mathbf{K}_{f} - \mathbf{k}),
\end{array}
\label{eq.2.7.5}
\end{equation}
where
\begin{equation}
\begin{array}{lcl}
  p_{fi} & = &
  \displaystyle\frac{1}{5}\;
  \Bigl\{
    c_{A}
    \displaystyle\sum\limits_{\alpha=1,2} \mathbf{e}^{(\alpha),*}\;
      J_{1} (\mathbf{k})\; -
    e^{- a^{2} k^{2}/4}\; c_{\rm p}
    \displaystyle\sum\limits_{\alpha=1,2} \mathbf{e}^{(\alpha),*}\;
      J_{2} (\mathbf{k})
  \Bigr\}.
\end{array}
\label{eq.2.7.6}
\end{equation}

For calculation of these integrals, we apply multipole expansion of wave function of photons.
According to formalism in , at quantum numbers $l_{i}=0$, $l_{f}=1$ and $l_{\rm ph}=1$ we have the following formula:
\begin{equation}
\begin{array}{l}
  \displaystyle\sum\limits_{\alpha=1,2}
    \mathbf{e}^{(\alpha)}\,
    \Bigl< k_{f} \Bigl|\, e^{-i\,\mathbf{kr}}\: \nabla\, \Bigr| \,k_{i} \Bigr>_{\mathbf{r}}\, =\,
  c_{M} \cdot J\,(1, 1) + c_{E1} \cdot J\,(1, 0) + c_{E2} \cdot J\,(1, 2)
\end{array}
\label{eq.2.7.7}
\end{equation}
where
\begin{equation}
\begin{array}{lcl}
  J\,(l_{f},n) & = &
  \displaystyle\int\limits^{+\infty}_{0}
    \displaystyle\frac{dR_{i}(r)}{dr}\:
    R^{*}_{f}(l,r)\,
    j_{n}(kr)\; r^{2} dr, \\

\end{array}
\label{eq.2.7.8}
\end{equation}
and
\begin{equation}
\begin{array}{lll}
  c_{M}\, = \,
    i\, \sqrt{\displaystyle\frac{3\pi}{2}}
    \displaystyle\sum\limits_{\mu=\pm 1}
      h_{\mu}\, \mu\: I\,(1, 1, 1, \mu), &

  c_{E1}\, =\,
    \sqrt{\pi}
    \displaystyle\sum\limits_{\mu=\pm 1}
      h_{\mu}\: I\,(1, 1, 0,\mu), &

  c_{E2}\, =\,
    - \, \sqrt{\displaystyle\frac{\pi}{2}}
    \displaystyle\sum\limits_{\mu=\pm 1}
      h_{\mu}\: I\,(1, 1, 2,\mu).
\end{array}
\label{eq.2.7.9}
\end{equation}
Here,
$I\,(l_{f}, l_{\rm ph}, n, \mu)$ are angular integrals defined in~\cite{Maydanyuk.2012.PRC,Maydanyuk_Zhang.2015.PRC}.
In the approximation of the leading integrals (in calculations, $J(1, 0)$ is the largest almost always, about on 10 times than other integrals),
we obtain:
\begin{equation}
\begin{array}{l}
  \displaystyle\sum\limits_{\alpha=1,2}
    \mathbf{e}^{(\alpha)}\,
    \Bigl< k_{f} \Bigl|\, e^{-i\,\mathbf{kr}}\: \nabla\, \Bigr| \,k_{i} \Bigr>_{\mathbf{r}}\, =\,
  c_{E1} \cdot J\,(1, 0).
\end{array}
\label{eq.2.7.10}
\end{equation}

We define the cross-section of the emitted photons on the bass of matrix element (\ref{eq.2.7.5}) in frameworks of formalism given in
\cite{Maydanyuk.2012.PRC,Maydanyuk_Zhang.2015.PRC} and we do not repeat it in this paper.
In result, we obtain the bremsstrahlung cross-section as
\begin{equation}
\begin{array}{ccl}
  \displaystyle\frac{d^{2}\,\sigma (\theta_{f})}{dw_{\rm ph}\,
  d\cos{\theta_{f}}} & = &
    \displaystyle\frac{e^{2}}{2\pi\,c^{5}}\:
      \displaystyle\frac{w_{\rm ph}\,E_{i}}{m_{\rm p}^{2}\,k_{i}} \;
      \biggl\{ p_{fi}\, \displaystyle\frac{d\, p_{fi}^{*}(\theta_{f})}{d\,\cos{\theta_{f}}}
      + {\rm c. c.} \biggr\},
\end{array}
\label{eq.2.7.11}
\end{equation}
where c.~c. is complex conjugation,
$p_{fi}$ is proportional to the electrical component $p_{\rm el}$ in Eqs.~(10) in \cite{Maydanyuk.2012.PRC}
and $d\,p_{fi} (\theta_{f})\, / d\,\cos{\theta_{f}}$ is defined by the same way as $d\,p\, (k_{i}, k_{f}, \theta_{f})\, / d\,\cos{\theta_{f}}$ in Ref.~\cite{Maydanyuk.2012.PRC}.

Experimental data of W\"{o}lfli, Hall and M\"{u}ller \cite{Wolfli.1971.PRL} provides information about the emitted photons in the coplanar geometry.
So, in order to compare our approach and calculations with those data,
we have to apply kinematic relations given by Eqs.~(2)--(3) in Ref.~\cite{Liu.1990.PRC.v42}
(see also Eqs.~(15) and (16) in \cite{Liu.1990.PRC.v41},
also \cite{Baye.1985.NPA}):
%
\begin{equation}
\begin{array}{ccl}
  E_{\rm ph} =
  E_{\rm p}\,
  \Bigl\{
    1 -
    \displaystyle\frac{4\, \sin^{2}\theta_{\alpha} + \sin^{2}\theta_{\rm p}}
      {4\, \sin^{2} (\theta_{\alpha}) + \theta_{\rm p}}
  \Bigr\}, &
  E_{f} = E_{i} - E_{\rm ph},
\end{array}
\label{eq.2.8.1}
\end{equation}
where $E_{\rm p}$ is incident proton energy in the laboratory frame,
$E_{\rm ph}$ is photon energy,
$E_{i}$ and $E_{f}$ are relative energies in the system-of-mass frame of the proton -- $\alpha$-particle system in the states before emission of photon (i.e. initial state) and after this emission (i.e. final state).
For the energy of the system in the initial state (i.e. before the emission of photon) in the center-of-mass frame we have $E_{i} = m_{\alpha}/(m_{\rm p} + m_{\alpha})\; E_{\rm p}$.
In the experimental setup \cite{Wolfli.1971.PRL} the coplanar arrangement of detectors was used. Here, protons are measured at the laboratory angle of $70^{\circ}$ concerning the beam axis (i.e. $\theta_{\rm p} = 70^{\circ}$), while the $\alpha$-particles are measured at opposite angle of $30^{\circ}$ ($\theta_{\alpha} = 30^{\circ}$).

For the scattering states, we use the space wave function $\varphi_{k}(\rhobf)$ in the spherically symmetric approximation
(i.e. where state of the full system is dependent only on relative distance $\rho = |\rhobf|$
between the scattered proton and center-of mass-of the $\alpha$-particle).
We have
\begin{equation}
\begin{array}{lcl}
  \varphi (\rhobf) =
  \varphi (\rho, \theta, \phi)
  & = &
  \displaystyle\frac{R_{l} (\rho)}{\rho}\:
  Y_{lm} (\theta, \phi),
\end{array}
\label{eq.2.8.2}
\end{equation}
where $R_{l} (\rho)$ and $Y_{lm} (\theta, \phi)$ are radial and angular wave functions,
$l$ and $m$ are quantum numbers.
General solution of the wave function we find as linear combination of two independent partial solutions
$c_{1l} (\rho)$ and $c_{2l} (\rho)$ as
\begin{equation}
\begin{array}{lcl}
  R_{l} (\rho)
  & = &
  A_{l}\, c_{1l} (\rho) + B_{l}\, c_{2l} (\rho),
\end{array}
\label{eq.2.8.3}
\end{equation}
where $A_{l}$ and $B_{l}$ are unknown amplitudes.
In the asymptotic limit, where we have only action of Coulomb forces, we apply
\begin{equation}
\begin{array}{lcl}
  c_{1l} (\rho) \to G_{l} (\rho),
  &
  c_{2l} (\rho) \to F_{l} (\rho),
\end{array}
\label{eq.2.8.4}
\end{equation}
where $G_{l}(\rho)$ and $F_{l}(\rho)$ are Coulomb wave functions.
Also quantum mechanics requires to apply condition of finite value of wave function $R_{l} (\rho)$ at zero $\rho=0$.
These conditions and normalization of the full wave function (for the scattering states in the continuous energy spectrum) allow us to determine the amplitudes
$A_{l}$ and $B_{l}$.
Using determination of our wave functions above, we can find phase shifts $\delta_{l}$, analyzed in details in~\cite{Liu.1990.PRC.v42}.
For this, we use definition (9) in \cite{Liu.1990.PRC.v41},
and obtain:
\begin{equation}
\begin{array}{lcl}
  \tan \delta_{l}
  & = &
  \displaystyle\frac{A_{l}}{B_{l}},
\end{array}
\label{eq.2.8.5}
\end{equation}
The radial wave functions $c_{1l}$ and $c_{2l}$ we calculate numerically. For description of interactions between proton and nucleus we extrapolate
potential with parameters from Ref.~\cite{Becchetti.1969.PR},
which was intensively studied and tested in wide region of nuclei for long time%
\footnote{In order to understand better a role of elastic processes in the scattering (where there is no formation of the compound $^{5}{\rm Li}$ nucleus),
we use only a real part of the potential~\cite{Becchetti.1969.PR} in calculations and obtaining results in Fig.~\ref{fig.2}.
Quantum number of full momentum $j$, spin projection of the scattering state are determined via the corresponding term in this potential
(so, the radial wave functions $R_{l}(\rho)$ are different for different such numbers,
see Ref.~\cite{Becchetti.1969.PR} for details).
As a next step, possible formation of the short-lived $^{5}{\rm Li}$ nucleus during the scattering can considered as inelastic process
and is taken into account via inclusion of the fusion amplitudes in Eq.~(\ref{eq.3.1}) into calculations and in obtaining results in Fig.~\ref{fig.3}.}.
In order to provide a numerical basis for analysis of experimental data via computer,
we shall use the following functions of errors~\cite{Maydanyuk_Zhang.2015.NPA}:
\begin{equation}
\begin{array}{lcl}
  \varepsilon_{1} & = &
  \displaystyle\frac{1}{N}
  \displaystyle\sum\limits_{k=1}^{N}
    \displaystyle\frac{\Bigl|\sigma^{\rm (theor)} (E_{k}) - \sigma^{\rm (exp)} (E_{k}) \Bigr|}{\sigma^{\rm (exp)} (E_{N})}, \\
  \varepsilon_{2} & = &
  \displaystyle\frac{1}{N}
  \displaystyle\sum\limits_{k=1}^{N}
    \displaystyle\frac{\Bigl|\sigma^{\rm (theor)} (E_{k}) - \sigma^{\rm (exp)} (E_{k}) \Bigr|}{\sigma^{\rm (exp)} (E_{k})}, \\
  \varepsilon_{3} & = &
  \displaystyle\frac{1}{N}
  \displaystyle\sum\limits_{k=1}^{N}
    \displaystyle\frac{\Bigl|\ln(\sigma^{\rm (theor)} (E_{k})) - \ln(\sigma^{\rm (exp)} (E_{k})) \Bigr|}{\ln(\sigma^{\rm (exp)} (E_{k}))},
\end{array}
\label{eq.2.8.6}
\end{equation}
where $\sigma^{\rm (theor)} (E_{k})$ and $\sigma^{\rm (exp)} (E_{k})$ are theoretical and experimental bremsstrahlung cross-sections at the proton incident energy $E_{k}$, the summation is performed over experimental data set,
and $N$ is number of experimental data points
($N=6$ for data~\cite{Wolfli.1971.PRL})%
\footnote{We introduced three different functions of errors in Eqs.~(\ref{eq.2.8.6}) in order to analyze the spectra more carefully in the different energy regions,
as they usually have exponential forms (for example, see Fig.~\ref{fig.2}(b)).}.



\section{Calculations, analysis
\label{sec.results}}

\begin{figure}[htbp]
\centerline{\includegraphics[width=97mm]{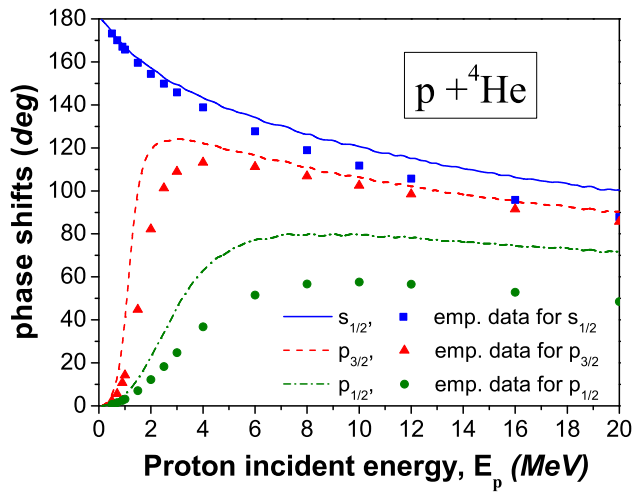}}
\vspace{-6mm}
\caption{\small (Color online)
Phase shifts for the scattering of protons off $\alpha$-particles calculated by Eq.~(\ref{eq.2.8.5}) in frameworks of our model
inside the incident proton energies till 20.0~MeV.
Here,
blue solid line is the calculated shift for $s_{1/2}$ state,
red dashed line is the calculated shift for $p_{3/2}$ state,
green dash-dotted line is the calculated shift for $p_{1/2}$ state,
blue squared points, red triangular points, and green circular points are
empirical data for the $s_{1/2}$, $p_{3/2}$  and $p_{1/2}$ states taken from Table III in~\cite{Arndt.1971.PRC}.
One can see that these calculations qualitatively corresponds to data calculated by Liu, Tang and Kanada~\cite{Liu.1990.PRC.v42} with potential set III of~\cite{Reichstein.1970.NPA} and empirical data of Arndt, Roper, and Shotwell~\cite{Arndt.1971.PRC}
(see Fig.~1 in~\cite{Liu.1990.PRC.v42}, for details).
This result can be considered as some good indication of logical coincidence of the wave functions for the scattering states in our model and the corresponding wave functions in the model~\cite{Liu.1990.PRC.v42}.
As a nest step, one can obtain essentially higher coincidence between our phase shifts and empirical data~\cite{Arndt.1971.PRC} that is subject of inverse theory (i.e., can be resolved completely, in principle) and, so, we should like to do not consider this technical task in this paper.
\label{fig.1}}
\end{figure}

We applied the method above to calculate the spectrum of the bremsstrahlung photons emitted during the scattering of protons off $\alpha$-particles at the incident proton energies up to 20~MeV.
Previous results given in Ref.~\cite{Liu.1990.PRC.v42} for the proton incident energies up to 25~MeV are based on the calculations of the corresponding wave functions for the scattering states, and maxima in the bremsstrahlung spectra are explained by phase shift of these wave functions.
By such a reason, we started our calculations from analysis of the phase shifts of the wave functions for the scattering states, which our approach provides.
Results of such calculations for the first some states are presented in Fig.~\ref{fig.1}.
Comparing these calculation with
data calculated by Liu, Tang and Kanada~\cite{Liu.1990.PRC.v42} with potential set III of~\cite{Reichstein.1970.NPA} and
empirical data of Arndt, Roper, and Shotwell~\cite{Arndt.1971.PRC},
we conclude that our calculated wave functions (for any states) have the same behavior, and are in some not bad agreement
(also see Fig. 1 in~\cite{Liu.1990.PRC.v42} for comparison).
Supposing, that better agreement between our calculations of the phase shifts and others results can be achieved using inverse scattering theory and is independent more technical task,
in this paper we should like to focus on the physical insight of what our approach would give in analysis of the bremsstrahlung experimental data.

\begin{figure}[htbp]
\centerline{\includegraphics[width=97mm]{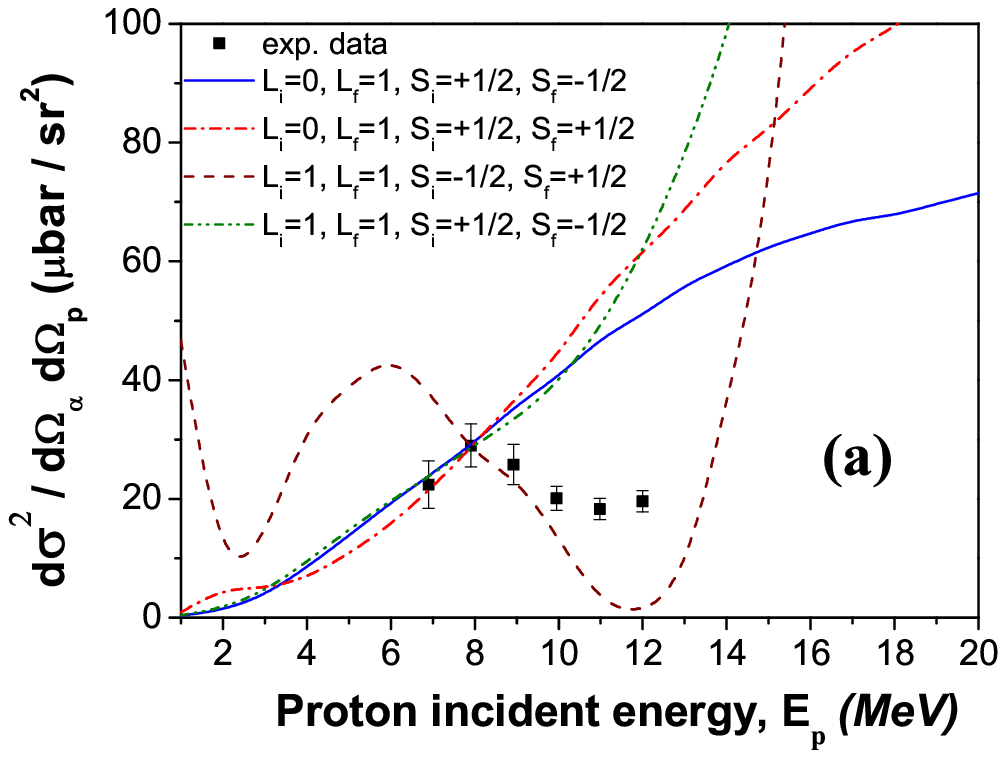}
\hspace{-9.5mm}\includegraphics[width=100mm]{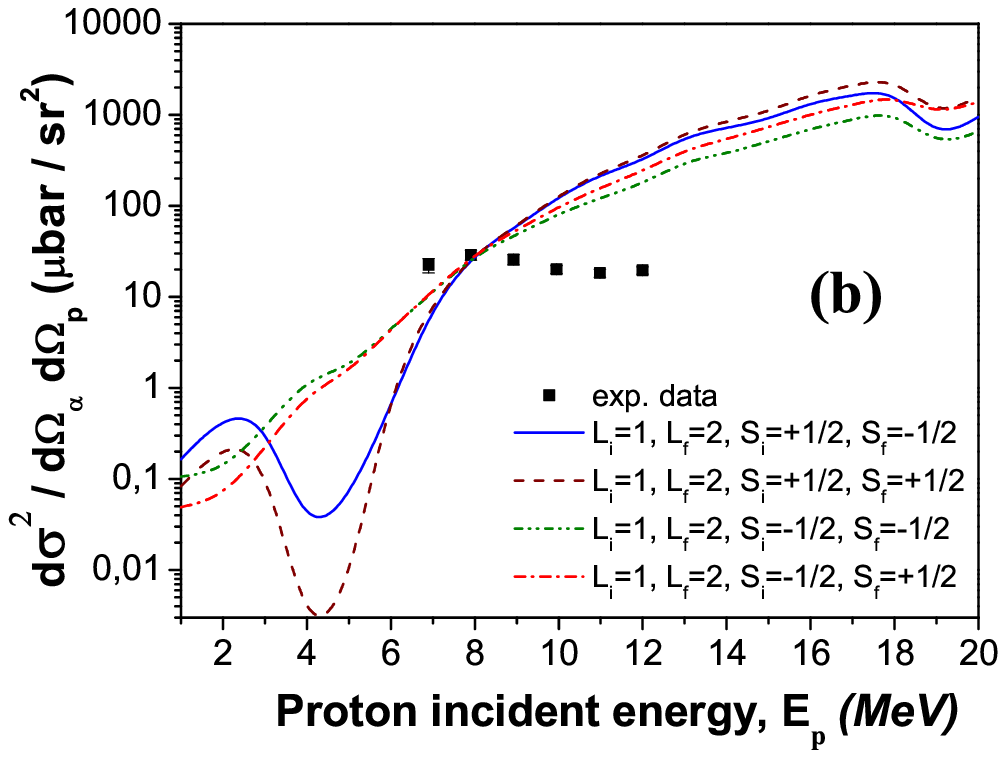}}
\vspace{-6mm}
\caption{\small (Color online)
Contributions of the emitted photons for some first transitions to the full bremsstrahlung spectrum in frameworks of our model
inside the incident proton energies from 1.0 to 20.0~MeV, and
experimental data (squared points) of W\"{o}lfli, Hall and M\"{u}ller \cite{Wolfli.1971.PRL}
(we normalize all calculated curves on the second point of experimental data).
%
(a) The calculated contributions for the transitions between states with $L_{i}=0,1$ and $L_{f}=1$.
Here, blue solid line is for $S_{i}=+1/2$ and $S_{f}=-1/2$,
red dash-dotted line is for $S_{i}=+1/2$ and $S_{f}=+1/2$,
brown dashed line is for $S_{i}=-1/2$ and $S_{f}=+1/2$,
green dash-double dotted line is for $S_{i}=+1/2$ and $S_{f}=-1/2$.
One can see that hump in the experimental cross-sections can be described if contribution formed in transition
between state $p_{i, 1/2}$ ($L_{i}=1$ and $S_{i}=-1/2$) and state $p_{f, 3/2}$ ($L_{f}=1$, $S_{f}=+1/2$) is leading.
%
(b) The calculated contributions for the transitions between states with $L_{i}=1$ and $L_{f}=2$.
Here, blue solid line is for $S_{i}=+1/2$ and $S_{f}=-1/2$,
red dash-dotted line is for $S_{i}=-1/2$ and $S_{f}=+1/2$,
brown dashed line is for $S_{i}=+1/2$ and $S_{f}=+1/2$,
green dash-double dotted line is for $S_{i}=-1/2$ and $S_{f}=-1/2$.
\label{fig.2}}
\end{figure}

As a next step, we calculated contributions of the emitted bremsstrahlung photons from some first transitions to the full spectrum in frameworks of our model.
Such calculations compared with experimental data  of W\"{o}lfli, Hall and M\"{u}ller \cite{Wolfli.1971.PRL} (we take them from Table~1 in that paper)
are presented in Fig.~\ref{fig.2}.
Here, one can see that the spectra are increased monotonously with increasing of the energy of proton for almost all studied transitions.
But we observe a visible maximum at the proton energy of $E_{\rm p} = 6$~MeV in the spectrum for the transition between states $p_{i, 1/2}$ ($L_{i}=1$, $S_{i}=-1/2$) and $p_{f, 3/2}$ ($L_{f}=1$, $S_{f}=+1/2$).
Also there are little variations in the spectra for the transitions between states with $L_{i}=1$ and $L_{f}=2$ at low proton energies (which can be related with limits in accurate calculations for the smaller cross-sections).

\begin{figure}[htbp]
\centerline{\includegraphics[width=97mm]{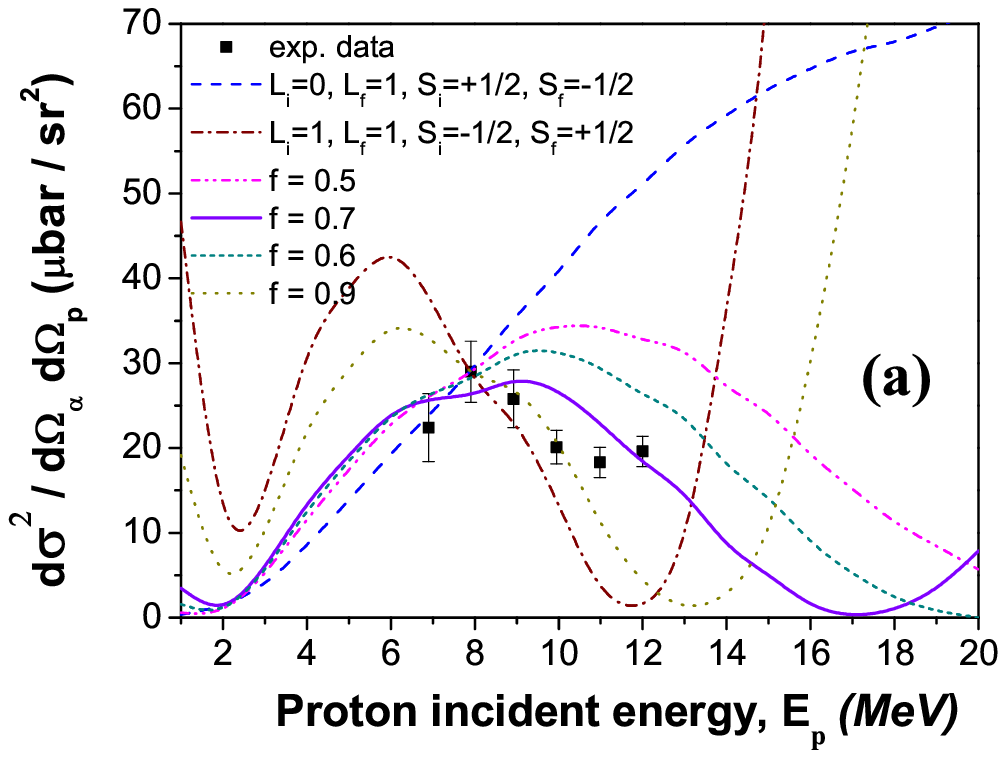}
\hspace{-9.5mm}\includegraphics[width=100mm]{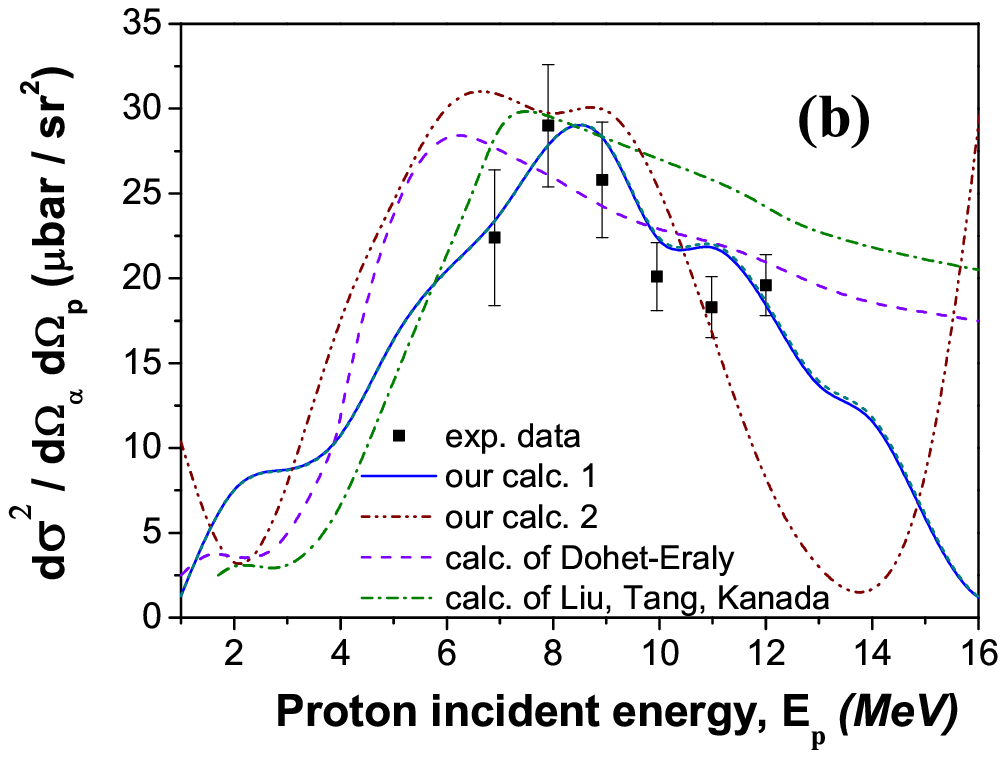}}
\vspace{-6mm}
\caption{\small (Color online)
The coplanar calculated cross sections of the bremsstrahlung photons emitted during the scattering of protons off the $\alpha$ particles, and
experimental data (squared points) of W\"{o}lfli, Hall and M\"{u}ller \cite{Wolfli.1971.PRL}.
(a) In this picture we demonstrate how one can displace hump of the full spectrum along the proton incident energy axis,
by changing ratio (defined by factor $f$) between
contribution for transition $p_{i, 1/2} \to p_{f, 3/2}$ and contribution for transition $s_{i, 1/2} \to p_{f, 1/2}$.
%
(b) Here,
blue solid line (calc. 1) is our calculated full spectrum based on all transitions given in Fig.~\ref{fig.2}(a, b) with amplitudes given in Eq.~(\ref{eq.3.2})
(such data are obtained on the basis of minimization of $\varepsilon_{2}$, we obtain $\varepsilon_{2} = 0.141$)
brown dash-double dotted line (calc. 2) is our calculated spectrum based on transitions given in Fig.~\ref{fig.2}(a) only,
green dash-dotted line is the spectrum calculated by Liu, Tang and Kanada in \cite{Liu.1990.PRC.v42} (see Fig. 2 in that paper),
violet dashed line is the spectrum calculated by Dohet-Eraly with used non-Siegert operator in \cite{Dohet-Eraly.2014.PRC.v89} (see Fig. 3(a) in that paper).
Our calculated curves for full spectrum have oscillatory behavior,
but they can be and some straighten
(and agreement with experimental data can be improved)
after using complex values for the amplitudes $f$ and inclusion of the other scattering states.
\label{fig.3}}
\end{figure}

The full bremsstrahlung spectrum is obtained at summation of the different contributions.
Different ratios between contributions give different shapes of the resulting spectrum.
In particular, one can find that hump in the experimental cross-sections \cite{Wolfli.1971.PRL} can be described if contribution formed in transition
$p_{i, 1/2} \to p_{f, 3/2}$ is leading.
Moreover, changing ratios between contributions, one can displace this hump along the proton incident energy axis.
For example, in Fig.~\ref{fig.3}(a) we demonstrate shift of such hump of the summarized spectrum, by changing ratio (defined by factor $f$) between contribution for transition $p_{i, 1/2} \to p_{f, 3/2}$ and contribution for transition $s_{i, 1/2} \to p_{f, 1/2}$.

Now we suppose that the wave function of the scattering state of proton off the $\alpha$-particle should include also possibility to form a combined system of $^{5}{\rm Li}$ which lives for some short time. In formation of such a combined system the fusion and opposite disintegration processes play important role.
For example, a clear picture of importance to study fusion
in forming composed systems
the $\alpha$-decay provide us,
which can be analyzed in three different stages:
(1) formation of a quantum object ($\alpha$- cluster having purely wave nature, behavior and quantum description) from some nucleons inside a space region of a parent nucleus,
(2) internal oscillations of this cluster before its further escaping outside, and
(3) tunneling transition of this cluster from nuclear region outside (related with tunneling through the barrier).
It is impossible to calculate correctly $\alpha$-decay half-lives for any nuclei without two first stages.
But, at the same time, calculations of the bremsstrahlung spectra in $\alpha$-decay, based only on inclusion of the third stage into the model, are very successful in description of experimental data (see
\cite{Maydanyuk.2006.EPJA,Maydanyuk.2008.EPJA,Giardina.2008.MPLA,Maydanyuk.2009.NPA,
Papenbrock.1998.PRLTA,Tkalya.1999.PHRVA,Jentschura.2008.PRC,Boie.2007.PRL}, and reference therein).

Here, we note our previous progress in quantum study of fusion processes in $\alpha$-capture by the $^{40}{\rm Ca}$ and $^{44}{\rm Ca}$ nuclei \cite{Maydanyuk_Zhang.2015.NPA}.
As we demonstrated in that paper,
quantum effects in fusion play important role (i.e they can be not small),
their inclusion into the model and calculations allows to essentially improve agreement with experimental data, and
quantum description of such processes can be performed via additional amplitudes (characterizing probabilities of presence of these processes in the different states).
Along with a logic in \cite{Maydanyuk_Zhang.2015.NPA}, instead of formula (\ref{eq.2.7.10})
we introduce a new one as
(here, we add numbers $l_{i}$, $s_{i}$, $s_{f}$ to designation of integral)
\begin{equation}
\begin{array}{l}
  \displaystyle\sum\limits_{\alpha=1,2}
    \mathbf{e}^{(\alpha)}\,
    \Bigl< k_{f} \Bigl|\, e^{-i\,\mathbf{kr}}\: \nabla\, \Bigr| \,k_{i} \Bigr>_{\mathbf{r}}\, =\,
  c_{E1}\,
  \displaystyle\sum\limits_{l_{i}, l_{f}, s_{i}, s_{f}}
  f_{l_{i}, s_{i}, l_{f}, s_{f}}\,
  J\,(l_{i}, s_{i}, l_{f}, s_{f}, n=l_{f}-1),
\end{array}
\label{eq.3.1}
\end{equation}
where $f_{l_{i}, s_{i}, l_{f}, s_{f}}$ are new real amplitudes characterized a possibility of formation of the compound $^{5}{\rm Li}$ nucleus and its breakup (disintegration) during the transition between the initial and final states
(further in the paper, \emph{amplitudes of the compound nucleus formation}, we have $0 \le f_{l_{i}, s_{i}, l_{f}, s_{f}} \le 1$).
Case of $f_{l_{i}=0, s_{i}=+1/2, l_{f}=1, s_{f}=\pm 1/2} = 1$ and
$f_{l_{i}, s_{i}, l_{f}, s_{f}} =0$ at other quantum numbers transforms formula (\ref{eq.3.1}) to the old Eq.~(\ref{eq.2.7.10}),
that corresponds to complete absence of the formation of the compound nucleus in the scattering
(i.e. the scattering takes place without any appearance of $^{5}{\rm Li}$).
So, we see that such coefficients characterize intensity of formation (fusion) and breakup (disintegration).

In order to clarify, if such processes are negligibly small or strong,
we shall look for values of the amplitudes $f_{l_{i}, s_{i}, l_{f}, s_{f}}$,
at which agreement between calculations and experimental data is the best.
In Fig.~\ref{fig.3}(b) we present results of such calculations in comparison with experimental data, where the found non-zero amplitudes are%
\footnote{All functions of errors $\varepsilon_{1}$, $\varepsilon_{2}$ and $\varepsilon_{3}$ give practically similar minimization results in the obtaining
amplitudes of compound nucleus formation. So, for simplicity of presentation, we use $\varepsilon_{2}$ for our calculations.}
\begin{equation}
\begin{array}{lllllllll}
  f_{0, +1/2, 1, -1/2} =
    f (s_{1/2} \to p_{1/2}) = 0.45263, &
  f_{1, -1/2, 2, -1/2} =
    f  (p_{1/2} \to d_{3/2}) = 0.655.
\end{array}
\label{eq.3.2}
\end{equation}
From analysis of the obtained calculations we conclude that
(1) inclusion of the amplitudes of the compound nucleus formation allows to essentially improve agreement between the full bremsstrahlung spectrum and the experimental data,
(2) transition $p_{1/2} \to p_{3/2}$ with a visible maximum at the proton energy of $E_{\rm p} = 6$~MeV in the spectrum  (see brown dashed line in Fig.~\ref{fig.2}(a)) does not play a role in description of hump in the experimental bremsstrahlung data.


\section{Conclusions
\label{sec.conclusions}}

In this paper we develop microscopic formalism of our bremsstrahlung model, which was previously successfully tested in description of experimental data in
alpha-decay, proton emission from nuclei, spontaneous fission, ternary fission, scattering of protons (at fixed incident energies) off nuclei in region of the emitted photons from lowest up to intermediate.
We focus on the scattering of protons off the $\alpha$-particles.
In description of the scattering states we implement our quantum formalism for calculations of wave functions based on the scattering theory.
In result, we obtain enough good agreement between the spectra calculated in frameworks of our approach and experimental data \cite{Wolfli.1971.PRL} (see Fig.~\ref{fig.3}(b)).
We formulate conclusions of application of this model in analysis of experimental data for this reaction.

\begin{enumerate}
\item
In the model connection between the bound states of nucleons inside the $\alpha$-particle, the scattering states and parameters of the emitted photon is obtained (see Eq.~(\ref{eq.2.7.6})).
But, influence of the parameters of the one-nucleon wave function on the bremsstrahlung spectrum is very small
(less than 1 percent).

\item

Analyzing bremsstrahlung experimental data, we observe compound nucleus $^{5}{\rm Li}$ formed in scattering of protons off the $\alpha$-particles.
Main transitions responsible for creation of such a nucleus are $s_{1/2} \to p_{1/2}$ and $p_{1/2} \to d_{3/2}$.

\item
In order to provide qualitative description of formation of the compound $^{5}{\rm Li}$ nucleus,
we introduce the new amplitudes
characterizing probabilities of formation of this nucleus and its breakup (disintegration) at definite transitions.
Analyzing experimental data, we obtain non-zero values
$f (s_{1/2} \to p_{1/2}) = 0.45263$ and $f  (p_{1/2} \to d_{3/2}) = 0.655$.


\end{enumerate}

\noindent
Note that only one experimental data point in~\cite{Wolfli.1971.PRL} at the emitted photon energy at~6.9~MeV gives a decreasing tendency of the spectrum at decreasing of energy of photons. This is not enough for a proper basis to conclude about such a behavior of the spectrum at the smaller photon energy.
If to compare the experimental data~\cite{Wolfli.1971.PRL} with the existed experimental data of the bremsstrahlung photons at
the scattering of protons off nuclei~\cite{Edington.1966.NP,Koehler.1967.PRL,Kwato_Njock.1988.PLB,Pinston.1989.PLB,
Pinston.1990.PLB,Clayton.1991.PhD,Clayton.1992.PRC.p1810,Clayton.1992.PRC.p1815,Chakrabarty.1999.PRC,Goethem.2002.PRL}
(see also reviews on situation in such a research \cite{Kamanin.1989.PEPAN,Pluiko.1987.PEPAN}),
$\alpha$-decay~\cite{Boie.2007.PRL,Boie.2009.PhD,Maydanyuk.2008.EPJA,Giardina.2008.MPLA,D'Arrigo.1994.PHLTA,Kasagi.1997.JPHGB,Kasagi.1997.PRLTA},
heavy-ion collisions~\cite{Bertholet.1987.NPA},
spontaneous fission~\cite{Ploeg.1995.PRC,Kasagi.1989.JPSJ,Luke.1991.PRC,Hofman.1993.PRC,Varlachev.2007.BRASP,Eremin.2010.IJMPE,Pandit.2010.PLB},
neutron-induced fission~\cite{Varlachev.2005.JETPL},
%
%
one can find that experimental information from~\cite{Wolfli.1971.PRL} is essentially not so rich.
In such regards, we see a sense in re-measuring of the bremsstrahlung emission of photons at the scattering of protons off the $\alpha$-particles inside the studied region of the emitted photon energy,
and we propose for experimental people to organize such experiments.


\section*{Acknowledgements
\label{sec.acknowledgements}}

S.~P.~M. thanks Institute of Modern Physics of Chinese Academy of Sciences for its warm hospitality and support.
This work was supported by the Major State Basic Research Development Program in China (No. 2015CB856903),
the National Natural Science Foundation of China (Grant Nos. 11575254 and 11175215), and
the Chinese Academy of Sciences fellowships for researchers from developing countries (No. 2014FFJA0003).


\appendix
\section{Summation over spin-isospin states
\label{sec.app.1}}

At first, we shall find summations in the term $M_{22}$ over spin and isospin states.
We shall study a case where nucleus is composed from even number $Z$ of protons and even number $N$ of neutrons.
For simplicity, we shall analyze a case of the scattering proton with spin -1/2 in this paper.
Write summations over proton and neutrons states separately as
\begin{equation}
\begin{array}{lcl}
\vspace{2mm}
  \displaystyle\sum\limits_{k=1}^{A+1}
  \displaystyle\sum\limits_{m=1, m \ne k}^{A+1}
    S_{km}\; =

  \displaystyle\sum\limits_{k=1}^{Z+1}
  \displaystyle\sum\limits_{m=1, m \ne k}^{Z+1}
    \delta_{\tau_{k}, 1/2}\: \delta_{\tau_{m}, 1/2}\; S_{km}\; +
  \displaystyle\sum\limits_{k=1}^{Z+1}
  \displaystyle\sum\limits_{m=1, m \ne k}^{N}
    \delta_{\tau_{k}, 1/2}\: \delta_{\tau_{m}, -1/2}\; S_{km}\; + \\
  + \quad
  \displaystyle\sum\limits_{k=1}^{N}
  \displaystyle\sum\limits_{m=1, m \ne k}^{Z+1}
    \delta_{\tau_{k}, -1/2}\: \delta_{\tau_{m}, 1/2}\; S_{km}\; +
  \displaystyle\sum\limits_{k=1}^{N}
  \displaystyle\sum\limits_{m=1, m \ne k}^{N}
    \delta_{\tau_{k}, -1/2}\: \delta_{\tau_{m}, -1/2}\; S_{km}.
\end{array}
\label{eq.2.5.1}
\end{equation}
Here, we consider the first term where we select summations over spin states:
\begin{equation}
\begin{array}{lcl}
\vspace{2mm}
  \displaystyle\sum\limits_{k=1}^{Z+1}
  \displaystyle\sum\limits_{m=1, m \ne k}^{Z+1}
    \delta_{\tau_{k}, 1/2}\: \delta_{\tau_{m}, 1/2}\; S_{km}\; = \\

\quad = \quad
  \displaystyle\sum\limits_{k=1}^{Z/2+1}
  \displaystyle\sum\limits_{m=1, m \ne k}^{Z/2+1}
    \delta_{\tau_{k}, 1/2}\: \delta_{\tau_{m}, 1/2}\;
    \delta_{\sigma_{k}, -1/2}\: \delta_{\sigma_{m}, -1/2}\; S_{km}\; +
  \displaystyle\sum\limits_{k=1}^{Z/2+1}
  \displaystyle\sum\limits_{m=1, m \ne k}^{Z/2}
    \delta_{\tau_{k}, 1/2}\: \delta_{\tau_{m}, 1/2}\;
    \delta_{\sigma_{k}, -1/2}\: \delta_{\sigma_{m}, 1/2}\; S_{km}\; + \\

\quad + \quad
  \displaystyle\sum\limits_{k=1}^{Z/2}
  \displaystyle\sum\limits_{m=1, m \ne k}^{Z/2+1}
    \delta_{\tau_{k}, 1/2}\: \delta_{\tau_{m}, 1/2}\;
    \delta_{\sigma_{k}, 1/2}\: \delta_{\sigma_{m}, -1/2}\; S_{km}\; +
  \displaystyle\sum\limits_{k=1}^{Z/2}
  \displaystyle\sum\limits_{m=1, m \ne k}^{Z/2}
    \delta_{\tau_{k}, 1/2}\: \delta_{\tau_{m}, 1/2}\;
    \delta_{\sigma_{k}, 1/2}\: \delta_{\sigma_{m}, 1/2}\; S_{km}.
\end{array}
\label{eq.2.5.2}
\end{equation}
Now we take into account that operator in (\ref{eq.2.4.7}) does not act on spin and isospin states.
So, we use properties of orthogonality of the wave functions in spin and isospin states.
For spin states we have
\begin{equation}
\begin{array}{lcl}
  \langle\, \downarrow_{i} | \downarrow_{j} \,\rangle =
  \langle\, \uparrow_{i} | \uparrow_{j} \,\rangle =
  \delta_{ij}, &

  \langle\, \downarrow_{i} | \uparrow_{j} \,\rangle =
  \langle\, \uparrow_{i} | \downarrow_{j} \,\rangle =
  0.
\end{array}
\label{eq.2.5.3}
\end{equation}
Analogical formulas we have for isospin states.
At first, we calculate terms for proton states in matrix element $M_{22}$:
\begin{equation}
\begin{array}{lcl}
  \displaystyle\sum\limits_{k=1}^{Z+1}
  \displaystyle\sum\limits_{m=1, m \ne k}^{Z+1}
    \biggl\{
    \Bigr\langle \psi_{k}(i_{\rm p}) \Bigl|\,
      e^{i\, c_{\rm p} \mathbf{kr} }\: \mathbf{p}
    \Bigr|\, \psi_{k}(i_{\rm p}) \Bigr\rangle\;
    \Bigr\langle \psi_{m}(j_{A}) \Bigl|\,
      f_{j}\, (\rhobf_{A j})
    \Bigr|\, \psi_{m}(j_{A}) \Bigr\rangle\; - \\
\vspace{4mm}
   - \quad
    \Bigr\langle \psi_{k}(i_{\rm p}) \Bigl|\,
      e^{i\, c_{\rm p} \mathbf{kr} }\: \mathbf{p}
    \Bigr|\, \psi_{m}(i_{\rm p}) \Bigr\rangle\;
    \Bigr\langle \psi_{m}(j_{A}) \Bigl|\,
      f_{j}\, (\rhobf_{A j})
    \Bigr|\, \psi_{k}(j_{A}) \Bigr\rangle\;
  \biggr\}\; = \\

  \vspace{1mm}
  =\quad
  4\; \displaystyle\sum\limits_{k=1}^{Z/2}
    \Bigr\langle \varphi_{k}(\mathbf{r}) \Bigl|\,
      e^{i\, c_{\rm p} \mathbf{kr} }\: \mathbf{p}
    \Bigr|\, \varphi_{k}(\mathbf{r}) \Bigr\rangle\;
  \displaystyle\sum\limits_{m=1, m \ne k}^{Z/2}
    \Bigr\langle \varphi_{m}(\rhobf_{A j}) \Bigl|\,
      f_{j}\, (\rhobf_{A j})
    \Bigr|\, \varphi_{m}(\rhobf_{A j}) \Bigr\rangle\; + \\

  \vspace{1mm}
  +\quad
  2\;
    \Bigr\langle \varphi_{k}(\mathbf{r}) \Bigl|\,
      e^{i\, c_{\rm p} \mathbf{kr} }\: \mathbf{p}
    \Bigr|\, \varphi_{k}(\mathbf{r}) \Bigr\rangle_{k=Z/2+1}\;
  \displaystyle\sum\limits_{m=1}^{Z/2}
    \Bigr\langle \varphi_{m}(\rhobf_{A j}) \Bigl|\,
      f_{j}\, (\rhobf_{A j})
    \Bigr|\, \varphi_{m}(\rhobf_{A j}) \Bigr\rangle + \\

  +\quad
  2\; \displaystyle\sum\limits_{k=1}^{Z/2}
    \Bigr\langle \varphi_{k}(\mathbf{r}) \Bigl|\,
      e^{i\, c_{\rm p} \mathbf{kr} }\: \mathbf{p}
    \Bigr|\, \varphi_{k}(\mathbf{r}) \Bigr\rangle\;
    \Bigr\langle \varphi_{m}(\rhobf_{A j}) \Bigl|\,
      f_{j}\, (\rhobf_{A j})
    \Bigr|\, \varphi_{m}(\rhobf_{A j}) \Bigr\rangle_{m=Z/2+1}.
\end{array}
\label{eq.2.5.4}
\end{equation}
%
Taking into account orthogonality conditions between isospin states for the one-nucleon wave functions, and
that operator $f_{j}\, (\rhobf_{A j})$ gives zero in acting on the one-nucleon wave function at neutron states,
we find other terms in matrix elements (\ref{eq.2.5.1}) to be equal to zero.
Summarizing, we find the term $M_{22}$:
\begin{equation}
\begin{array}{lcl}
\vspace{1mm}
  M_{22} & = &
  c_{\rm p}\,
    \displaystyle\sum\limits_{j=1}^{Z_{A}}\,
    \displaystyle\frac{1}{A\,(A+1)}\;
    \biggl\{

  4\; \displaystyle\sum\limits_{k=1}^{Z/2}
    \Bigr\langle \varphi_{k}(\mathbf{r}) \Bigl|\,
      e^{i\, c_{\rm p} \mathbf{kr} }\: \mathbf{p}
    \Bigr|\, \varphi_{k}(\mathbf{r}) \Bigr\rangle\;
  \displaystyle\sum\limits_{m=1, m \ne k}^{Z/2}
    \Bigr\langle \varphi_{m}(\rhobf_{A j}) \Bigl|\,
      f_{j}\, (\rhobf_{A j})
    \Bigr|\, \varphi_{m}(\rhobf_{A j}) \Bigr\rangle\; + \\

  \vspace{1mm}
  & + &
  2\;
    \Bigr\langle \varphi_{k}(\mathbf{r}) \Bigl|\,
      e^{i\, c_{\rm p} \mathbf{kr} }\: \mathbf{p}
    \Bigr|\, \varphi_{k}(\mathbf{r}) \Bigr\rangle_{k=Z/2+1}\;
  \displaystyle\sum\limits_{m=1}^{Z/2}
    \Bigr\langle \varphi_{m}(\rhobf_{A j}) \Bigl|\,
      f_{j}\, (\rhobf_{A j})
    \Bigr|\, \varphi_{m}(\rhobf_{A j}) \Bigr\rangle\; + \\

  & + &
  2\; \displaystyle\sum\limits_{k=1}^{Z/2}
    \Bigr\langle \varphi_{k}(\mathbf{r}) \Bigl|\,
      e^{i\, c_{\rm p} \mathbf{kr} }\: \mathbf{p}
    \Bigr|\, \varphi_{k}(\mathbf{r}) \Bigr\rangle\;
    \Bigr\langle \varphi_{m}(\rhobf_{A j}) \Bigl|\,
      f_{j}\, (\rhobf_{A j})
    \Bigr|\, \varphi_{m}(\rhobf_{A j}) \Bigr\rangle_{m=Z/2+1}
  \biggr\}.
\end{array}
\label{eq.2.5.5}
\end{equation}

Now we shall calculate the term $M_{21}$.
Let us consider such a summation where we write separately terms for different spin and isospin states
\begin{equation}
\begin{array}{lcl}
\vspace{2mm}
  \displaystyle\sum\limits_{k=1}^{A+1} S_{km}\; & = & 
  \displaystyle\sum\limits_{k=1}^{Z/2+1}
    \delta_{\tau_{k}, 1/2}\;
    \delta_{\sigma_{k}, -1/2}\; S_{km}\; +
  \displaystyle\sum\limits_{k=1}^{Z/2}
    \delta_{\tau_{k}, 1/2}\;
    \delta_{\sigma_{k}, 1/2}\; S_{km}\; + \\

  & + &
  \displaystyle\sum\limits_{k=1}^{N/2}
    \delta_{\tau_{k}, -1/2}\;
    \delta_{\sigma_{k}, -1/2}\; S_{km}\; +
  \displaystyle\sum\limits_{k=1}^{N/2}
    \delta_{\tau_{k}, -1/2}\;
    \delta_{\sigma_{k}, 1/2}\; S_{km}.
\end{array}
\label{eq.2.5.6}
\end{equation}
Taking into account orthogonality conditions between isospin states for the one-nucleon wave functions, and
that operator in matrix element $M_{21}$ gives zero in acting on the one-nucleon wave function at neutron states,
we obtain:
\begin{equation}
\begin{array}{lcl}
\vspace{0.5mm}
  M_{21} & = &
  \displaystyle\frac{c_{A}}{A+1}\;
  \biggl\{

  2\; \displaystyle\sum\limits_{k=1}^{Z/2}
    \Bigr\langle \varphi_{k}(\mathbf{r}) \Bigl|\,
      e^{i\, c_{\rm p} \mathbf{kr} }\,
      e^{-i \mathbf{k}\mathbf{r}}\,
      f_{\rm p} (\rhobf_{\rm p})\: \mathbf{p}
    \Bigr|\, \varphi_{k}(\mathbf{r}) \Bigr\rangle\; + \;
  2\; \displaystyle\sum\limits_{k=1}^{N/2}
    \Bigr\langle \varphi_{k}(\mathbf{r}) \Bigl|\,
      e^{i\, c_{\rm p} \mathbf{kr} }\,
      e^{-i \mathbf{k}\mathbf{r}}\,
      f_{\rm p} (\rhobf_{\rm p})\: \mathbf{p}
    \Bigr|\, \varphi_{k}(\mathbf{r}) \Bigr\rangle\; + \\
  & + &
  \Bigr\langle \varphi_{k}(\mathbf{r}) \Bigl|\,
      e^{i\, c_{\rm p} \mathbf{kr} }\,
      e^{-i \mathbf{k}\mathbf{r}}\,
      f_{\rm p} (\rhobf_{\rm p})\: \mathbf{p}
    \Bigr|\, \varphi_{k}(\mathbf{r}) \Bigr\rangle_{k=Z/2+1}
  \biggr\}.
\end{array}
\label{eq.2.5.7}
\end{equation}
In a case of the $\alpha$-particle (in the ground state) as the target-nucleus,  we calculate such integrals
(see Appendix~\ref{sec.app.2}):
\begin{equation}
\begin{array}{lcl}
  \Bigl\langle \varphi_{000}(\rhobf)\, \Bigl|\,
    e^{-i \mathbf{k} \rhobfsm}\,
  \Bigr|\, \varphi_{000}(\rhobf)\, \Bigr\rangle
  & = &
  e^{-\, (a^{2} k_{x}^{2} + b^{2} k_{y}^{2} + c^{2} k_{z}^{2})\,/4}, \\

  \Bigl\langle \varphi_{000}(\rhobf)\, \Bigl|\,
    e^{i\, c_{\rm p} \mathbf{k}\rhobfsm}\: \mathbf{p}
  \Bigr|\, \varphi_{000}(\rhobf)\, \Bigr\rangle
  & = &
  \displaystyle\frac{i\, c_{\rm p}\, \mathbf{k}}{2}\;
  e^{-\, c_{\rm p}^{2}\, (a^{2} k_{x}^{2} + b^{2} k_{y}^{2} + c^{2} k_{z}^{2})\,/4}, \\

  \Bigr\langle \varphi_{000}(\rhobf) \Bigl|\,
    e^{i\, c_{\rm p} \mathbf{k}\rhobfsm }\,
    e^{-i \mathbf{k}\rhobfsm}\: \mathbf{p}
  \Bigr|\, \varphi_{000}(\rhobf) \Bigr\rangle
  & = &
  \displaystyle\frac{i\, (c_{\rm p}-1)\, \mathbf{k}}{2}\;
  e^{-\, (c_{\rm p}-1)^{2}\, (a^{2} k_{x}^{2} + b^{2} k_{y}^{2} + c^{2} k_{z}^{2})\,/4}.
\end{array}
\label{eq.2.5.8}
\end{equation}

\section{Calculations of one-nucleon space matrix elements over bound states of nucleus
\label{sec.app.2}}

In this Appendix we shall find the matrix elements.
For the bound state (which we label by index $k_{\rm bound} = 1 \ldots Z/2$) we have
\begin{equation}
\begin{array}{lcl}
  \Bigr\langle \varphi_{k_{\rm bound}}(\mathbf{r}) \Bigl|\,
      e^{i\, c_{\rm p} \mathbf{kr} }\: \mathbf{p}
    \Bigr|\, \varphi_{k_{\rm bound}}(\mathbf{r}) \Bigr\rangle , &

  \Bigr\langle \varphi_{k_{\rm bound}}(\rhobf_{A j}) \Bigl|\,
      f_{j}\, (\rhobf_{A j})
    \Bigr|\, \varphi_{k_{\rm bound}}(\rhobf_{A j}) \Bigr\rangle. \\
\end{array}
\label{eq.app.0.1}
\end{equation}
For the scattering state (which we label by index $k_{\rm scat} = Z/2 + 1$) we have
\begin{equation}
\begin{array}{lcl}
  \Bigr\langle \varphi_{k_{\rm scat}}(\mathbf{r}) \Bigl|\,
      e^{i\, c_{\rm p} \mathbf{kr} }\: \mathbf{p}
    \Bigr|\, \varphi_{k_{\rm scat}}(\mathbf{r}) \Bigr\rangle , &

  \Bigr\langle \varphi_{k_{\rm scat}}(\rhobf_{A j}) \Bigl|\,
      f_{j}\, (\rhobf_{A j})
    \Bigr|\, \varphi_{k_{\rm scat}}(\rhobf_{A j}) \Bigr\rangle. \\
\end{array}
\label{eq.app.0.2}
\end{equation}
For the first term of the matrix element we have:
\begin{equation}
\begin{array}{lcl}
  \Bigr\langle \varphi_{k_{\rm bound}}(\mathbf{r}) \Bigl|\,
      e^{i\, c_{\rm p} \mathbf{kr} }\,
      e^{-i \mathbf{k}\mathbf{r}}\,
      f_{\rm p} (\rhobf_{\rm p})\: \mathbf{p}
    \Bigr|\, \varphi_{k_{\rm bound}}(\mathbf{r}) \Bigr\rangle, &

  \Bigr\langle \varphi_{k_{\rm scat}}(\mathbf{r}) \Bigl|\,
      e^{i\, c_{\rm p} \mathbf{kr} }\,
      e^{-i \mathbf{k}\mathbf{r}}\,
      f_{\rm p} (\rhobf_{\rm p})\: \mathbf{p}
    \Bigr|\, \varphi_{k_{\rm scat}}(\mathbf{r}) \Bigr\rangle,
\end{array}
\label{eq.app.0.3}
\end{equation}
where $f_{\rm p} (\rhobf_{\rm p})=1$ for proton.


\subsection{One-nucleon space wave function and matrix element
\label{sec.app.2.1}}

We shall choose the space wave function of one nucleon in the gaussian form
\begin{equation}
  \varphi_{n_{x},n_{y},n_{z}} (\mathbf{r}) =
  N_{x}\,N_{y}\,N_{z} \cdot
  \exp{\Bigl[-\,\displaystyle\frac{1}{2}\,\Bigl(\displaystyle\frac{x^{2}}{a^{2}} +
    \displaystyle\frac{y^{2}}{b^{2}} + \displaystyle\frac{z^{2}}{c^{2}}\Bigr) \Bigr]} \cdot
  H_{n_{x}} \Bigl(\displaystyle\frac{x}{a} \Bigr)\,
  H_{n_{y}} \Bigl(\displaystyle\frac{y}{b} \Bigr)\,
  H_{n_{z}} \Bigl(\displaystyle\frac{z}{c} \Bigr),
\label{eq.app.1.1.1}
\end{equation}
where $H_{n_{x}}$, $H_{n_{y}}$ and $H_{n_{z}}$ are the Hermitian polynomials,
$N_{x}$, $N_{y}$, $N_{z}$ are the normalized coefficients.
The unknown normalized coefficients are calculated from the normalization condition:
\begin{equation}
\begin{array}{lcl}
  \vspace{2mm}
  \displaystyle\int
    \Bigl| N_{x}\,
    \exp{\Bigl[-\,\displaystyle\frac{x^{2}}{2a^{2}} \Bigr]}\,
    H_{n_{x}} \Bigl(\displaystyle\frac{x}{a} \Bigr)\, \Bigl|^{2}\; dx = 1, & 

  \vspace{2mm}
  \displaystyle\int
    \Bigl| N_{y}\,
    \exp{\Bigl[-\,\displaystyle\frac{y^{2}}{2b^{2}} \Bigr]}\,
    H_{n_{y}} \Bigl(\displaystyle\frac{y}{b} \Bigr)\Bigl|^{2}\; dy = 1, & 
  \displaystyle\int
    \Bigl| N_{z}\,
    \exp{\Bigl[-\,\displaystyle\frac{z^{2}}{2c^{2}} \Bigr]}\,
    H_{n_{z}} \Bigl(\displaystyle\frac{z}{c} \Bigr) \Bigl|^{2}\; dz = 1.
\end{array}
\label{eq.app.1.1.2}
\end{equation}
Taking the properties of the Hermitian polynomials into account (see \cite{Landau.v3.1989}, p.~749):
\begin{equation}
\begin{array}{cccc}
  \displaystyle\int\limits_{-\infty}^{+\infty} e^{-x^{2}}\, H_{n}^{2}(x)\; dx = 2^{n}\, n!\, \sqrt{\pi}, &
  H_{0} = 1, &
  H_{1} = 2x, &
  H_{2} = 4x^{2} - 2,
\end{array}
\label{eq.app.1.1.3}
\end{equation}
we obtain:
\begin{equation}
\begin{array}{ccc}
  N_{x} = \displaystyle\frac{1}{\pi^{1/4} \sqrt{ a\, 2^{n_{x}}\, n_{x}! }}, &
  N_{y} = \displaystyle\frac{1}{\pi^{1/4} \sqrt{ b\, 2^{n_{y}}\, n_{y}! }}, &
  N_{z} = \displaystyle\frac{1}{\pi^{1/4} \sqrt{ c\, 2^{n_{z}}\, n_{z}! }}.
\end{array}
\label{eq.app.1.1.4}
\end{equation}

We shall calculate the second matrix element in eqs.~(\ref{eq.app.0.1}) for protons in form
\begin{equation}
\begin{array}{lcl}
  \Bigr\langle \varphi_{k_{\rm bound}}(\rhobf) \Bigl|\,
    f_{A j}\, (\rhobf)
  \Bigr|\, \varphi_{k_{\rm bound}}(\rhobf) \Bigr\rangle =

  \Bigr\langle \varphi_{k_{\rm bound}}(\rhobf) \Bigl|\,
    e^{-i \mathbf{k} \rhobfsm}\,
  \Bigr|\, \varphi_{k_{\rm bound}}(\rhobf) \Bigr\rangle,
\end{array}
\label{eq.app.1.1.5}
\end{equation}
where we take into account $z_{p}=1$ for protons.
%
Substituting form of the wave function into Eq.~(\ref{eq.app.1.1.1}), we calculate the matrix element for the $\alpha$ particle:
\begin{equation}
\begin{array}{lcl}
  & &
    \Bigl\langle \varphi_{n_{x},n_{y},n_{z}}(\rhobf)\, \Bigl|\,
      e^{-i \mathbf{k} \rhobfsm}\,
    \Bigr|\, \varphi_{n_{x},n_{y},n_{z}}(\rhobf)\, \Bigr\rangle =

  \displaystyle\int\:
    \varphi_{n_{x},n_{y},n_{z}}^{2} (\rhobf)\,
    e^{-i \mathbf{k} \rhobfsm}\; \rhobf\; =

  I_{x}(n_{x})\, I_{y}(n_{y})\, I_{z}(n_{z}),
\end{array}
\label{eq.app.1.1.6}
\end{equation}
where
\begin{equation}
\begin{array}{lcl}
\vspace{1mm}
  I_{x}
  & = &
  N_{\alpha,x}^{2}\;
    \exp{\Bigl[-\, a^{2} k_{x}^{2}/4 \Bigr]}\;  
  \displaystyle\int\:
    \exp{\Bigl[-\,\displaystyle\frac{(x + i\,a^{2} k_{x}/2)^{2} }{a^{2}} \Bigr]}\,
    H_{n_{x}}^{2} \Bigl(\displaystyle\frac{x}{a} \Bigr)\; dx.
\end{array}
\label{eq.app.1.1.7}
\end{equation}
and solutions for $I_{y} (n_{y})$ and $I_{z} (n_{z})$ are obtained after change of indexes $x \to y$ and $x \to z$.

\subsection{Case of the scattering of the proton off $\alpha$-particle
\label{sec.app.2.2}}

Now let us consider case when the $\alpha$ particle is in the ground state ($n_{x} = n_{y} = n_{z} = 0$). We have
$H_{n_{x}=0} = 1$,
$H_{n_{y}=0} = 1$,
$H_{n_{z}=0} = 1$.
In approximation, integral in eq.~(\ref{eq.app.1.1.7}) over complex variable $\tilde{x} = x + i\,a^{2} k_{x}/2$
has solution:
\begin{equation}
\begin{array}{lcl}
  \displaystyle\int\:
      \exp{\Bigl[-\,\displaystyle\frac{(x + i\,a^{2} k_{x}/2)^{2} }{a^{2}} \Bigr]}\;
    dx_{i} =
  N_{\alpha,x}^{-2}
\end{array}
\label{eq.app.1.2.1}
\end{equation}
and we obtain:
\begin{equation}
\begin{array}{lcl}
  I_{\alpha,x} (n_{x}=0)
  & = &
  \exp{\Bigl[-\, a^{2} k_{x}^{2}/4 \Bigr]}.
\end{array}
\label{eq.app.1.2.2}
\end{equation}
Now we calculate the matrix element (\ref{eq.app.1.1.6}):
\begin{equation}
\begin{array}{lcl}
  \Bigl\langle \varphi_{000}(\rhobf)\, \Bigl|\,
    e^{-i \mathbf{k} \rhobfsm}\,
  \Bigr|\, \varphi_{000}(\rhobf)\, \Bigr\rangle
  & = &
  e^{-\, (a^{2} k_{x}^{2} + b^{2} k_{y}^{2} + c^{2} k_{z}^{2})\,/4}.
\end{array}
\label{eq.app.1.2.3}
\end{equation}

We shall calculate the first matrix element (\ref{eq.app.0.1}) for protons in form
\begin{equation}
\begin{array}{lcl}
  \Bigr\langle \varphi_{k_{\rm bound}}(\rhobf) \Bigl|\,
    e^{i\, c_{\rm p} \mathbf{k\rhobfsm} }\: \mathbf{p}
  \Bigr|\, \varphi_{k_{\rm bound}}(\rhobf) \Bigr\rangle =

  - i\hbar\:
  \Bigr\langle \varphi_{k_{\rm bound}}(\rhobf) \Bigl|\,
    e^{i\, c_{\rm p} \mathbf{k\rhobfsm} }\: \displaystyle\frac{\mathbf{d}}{\mathbf{d}\rhobf}
  \Bigr|\, \varphi_{k_{\rm bound}}(\rhobf) \Bigr\rangle.
\end{array}
\label{eq.app.1.2.4}
\end{equation}
Consider a case of the $\alpha$-particle in the ground state.
We have:
\begin{equation}
\begin{array}{lcl}
  & &
  \Bigl\langle \varphi_{000}(\rhobf)\, \Bigl|\,
    e^{i\, c_{\rm p} \mathbf{k\rhobfsm} }\: \displaystyle\frac{\mathbf{d}}{\mathbf{d}\rhobf}
  \Bigr|\, \varphi_{000}(\rhobf)\, \Bigr\rangle =
  I_{2,x}(n_{x}=0)\, I_{2,y}(n_{y}=0)\, I_{2,z}(n_{z}=0),
\end{array}
\label{eq.app.1.2.5}
\end{equation}
where
\begin{equation}
\begin{array}{lcl}
  I_{2,x} (n_{x}, a) & = &
  N_{\alpha,x}^{2}
  \displaystyle\int
    e^{-\,\frac{x^{2}}{2a^{2}} }\,
    \displaystyle\frac{d}{dx}\;
    e^{-\,\frac{x^{2}}{2a^{2}} }\,
    e^{-ic_{\rm p}\, k_{x} x}\; dx
\end{array}
\label{eq.app.1.2.6}
\end{equation}
and solutions for $I_{y} (n_{y})$ and $I_{z} (n_{z})$ are obtained after change of indexes $x \to y$ and $x \to z$.
Calculate this integral:
\begin{equation}
\begin{array}{lcl}
  I_{2,x} (n_{x}, a) & = &
  -\, \,\frac{N_{\alpha,x}^{2}}{a^{2}}\,
  \displaystyle\int
    x\, e^{-\,\frac{x^{2}}{a^{2}} }\,
    e^{-ic_{\rm p}\, k_{x} x}\; dx.
\end{array}
\label{eq.app.1.2.7}
\end{equation}
We simplify this integral and obtain:
\begin{equation}
\begin{array}{lcl}
\vspace{1mm}
  I_{2,x}(n_{x}, a)
  & = &
  -\, \,\displaystyle\frac{N_{\alpha,x}^{2}}{a^{2}}\;
  \exp{\Bigl[ - c_{\rm p}^{2}\, a^{2} k_{x}^{2}/4 \Bigr]}\;
  \displaystyle\int
    (x + ic_{\rm p}\, a^{2}\, k_{x}/2)\,
    \exp{\Bigl[ -\,\displaystyle\frac{(x + ic_{\rm p}\, a^{2}\, k_{x}/2)^{2}}{a^{2}}
         \Bigr]}\; dx\; + \\

  & + &
  \displaystyle\frac{N_{\alpha,x}^{2}}{a^{2}}\;
  \exp{\Bigl[ - c_{\rm p}^{2}\, a^{2} k_{x}^{2}/4 \Bigr]}\;
    (ic_{\rm p}\, a^{2}\, k_{x}/2)\:
  \displaystyle\int
    \exp{\Bigl[ -\,\displaystyle\frac{(x + ic_{\rm p}\, a^{2}\, k_{x}/2)^{2}}{a^{2}}
         \Bigr]}\; dx.
\end{array}
\label{eq.app.1.2.8}
\end{equation}

For next integration of the first integral in this expression, we use property:
\begin{equation}
\begin{array}{lcl}
  \displaystyle\frac{d}{dx}
  \exp{\Bigl[ -\,\displaystyle\frac{(x + ic_{\rm p}\, a^{2}\, k_{x}/2)^{2}}{a^{2}} \Bigr]}\; =

  -\,\displaystyle\frac{2}{a^{2}}\:
  (x + ic_{\rm p}\, a^{2}\, k_{x}/2)\:
  \exp{\Bigl[ -\,\displaystyle\frac{(x + ic_{\rm p}\, a^{2}\, k_{x}/2)^{2}}{a^{2}} \Bigr]}.
\end{array}
\label{eq.app.1.2.9}
\end{equation}
and we write
\begin{equation}
\begin{array}{lcl}
  (x + ic_{\rm p}\, a^{2}\, k_{x}/2)\:
  \exp{\Bigl[ -\,\displaystyle\frac{(x + ic_{\rm p}\, a^{2}\, k_{x}/2)^{2}}{a^{2}} \Bigr]}\; =

  -\,\displaystyle\frac{a^{2}}{2}\:
  \displaystyle\frac{d}{dx}
  \exp{\Bigl[ -\,\displaystyle\frac{(x + ic_{\rm p}\, a^{2}\, k_{x}/2)^{2}}{a^{2}} \Bigr]}.
\end{array}
\label{eq.app.1.2.10}
\end{equation}
Taking this equation into account, we find the first term in eq.~(\ref{eq.app.1.2.8}) as
\begin{equation}
\begin{array}{lcl}
  -\, \,\displaystyle\frac{N_{\alpha,x}^{2}}{a^{2}}\;
  \exp{\Bigl[ - c_{\rm p}^{2}\, a^{2} k_{x}^{2}/4 \Bigr]}\;
  \displaystyle\int
    (x + ic_{\rm p}\, a^{2}\, k_{x}/2)\,
    \exp{\Bigl[ -\,\displaystyle\frac{(x + ic_{\rm p}\, a^{2}\, k_{x}/2)^{2}}{a^{2}}
         \Bigr]}\; dx\; = 0,
\end{array}
\label{eq.app.1.2.11}
\end{equation}
where we suppose
\begin{equation}
  \exp{\Bigl[ -\,\displaystyle\frac{(x + ic_{\rm p}\, a^{2}\, k_{x}/2)^{2}}{a^{2}}
    \Bigr]}^{x \to +\infty}_{x \to -\infty}\; =\; 0.
\label{eq.app.1.2.12}
\end{equation}
The second term in eq.~(\ref{eq.app.1.2.8}) can be expressed via the previous found one:
\begin{equation}
\begin{array}{lcl}
\vspace{1mm}
  & &
  \displaystyle\frac{N_{\alpha,x}^{2}}{a^{2}}\;
  \exp{\Bigl[ - c_{\rm p}^{2}\, a^{2} k_{x}^{2}/4 \Bigr]}\;
    (ic_{\rm p}\, a^{2}\, k_{x}/2)\:
  \displaystyle\int
    \exp{\Bigl[ -\,\displaystyle\frac{(x + ic_{\rm p}\, a^{2}\, k_{x}/2)^{2}}{a^{2}}
         \Bigr]}\; dx\; =

  \displaystyle\frac{ic_{\rm p}\, k_{x}}{2}\;
  \exp{\Bigl[ - c_{\rm p}^{2}\, a^{2} k_{x}^{2}/4 \Bigr]}.
\end{array}
\label{eq.app.1.2.13}
\end{equation}
So, we find solution for integral:
\begin{equation}
\begin{array}{lcl}
  I_{2,x}(n_{x}, a)
  & = &
  \displaystyle\frac{ic_{\rm p}\, k_{x}}{2}\;
  \exp{\Bigl[ - c_{\rm p}^{2}\, a^{2} k_{x}^{2}/4 \Bigr]}
\end{array}
\label{eq.app.1.2.14}
\end{equation}
and we calculate the matrix element (\ref{eq.app.0.2}) and  (\ref{eq.app.0.3}).
Now we rewrite the all obtained results:
\begin{equation}
\begin{array}{lcl}
  \Bigl\langle \varphi_{000}(\rhobf)\, \Bigl|\,
    e^{-i \mathbf{k} \rhobfsm}\,
  \Bigr|\, \varphi_{000}(\rhobf)\, \Bigr\rangle
  & = &
  e^{-\, (a^{2} k_{x}^{2} + b^{2} k_{y}^{2} + c^{2} k_{z}^{2})\,/4}, \\

  \Bigl\langle \varphi_{000}(\rhobf)\, \Bigl|\,
    e^{i\, c_{\rm p} \mathbf{k}\rhobfsm}\: \mathbf{p}
  \Bigr|\, \varphi_{000}(\rhobf)\, \Bigr\rangle
  & = &
  \displaystyle\frac{i\, c_{\rm p}\, \mathbf{k}}{2}\;
  e^{-\, c_{\rm p}^{2}\, (a^{2} k_{x}^{2} + b^{2} k_{y}^{2} + c^{2} k_{z}^{2})\,/4}, \\

  \Bigr\langle \varphi_{000}(\rhobf) \Bigl|\,
    e^{i\, c_{\rm p} \mathbf{k}\rhobfsm }\,
    e^{-i \mathbf{k}\rhobfsm}\: \mathbf{p}
  \Bigr|\, \varphi_{000}(\rhobf) \Bigr\rangle
  & = &
  \displaystyle\frac{i\, (c_{\rm p}-1)\, \mathbf{k}}{2}\;
  e^{-\, (c_{\rm p}-1)^{2}\, (a^{2} k_{x}^{2} + b^{2} k_{y}^{2} + c^{2} k_{z}^{2})\,/4}.
\end{array}
\label{eq.app.1.2.15}
\end{equation}




\vspace{5mm}
\begin{thebibliography}{99}
\bibitem{Liu.1981.PRC.v23}
  Q.~K.~K.~Liu, H.~Kanada, and Y.~C.~Tang,
\newblock
  \emph{Microscopic study of $^{3}{\rm He} (\alpha, \gamma) ^{7}{\rm Be}$ electric-dipole capture reaction},
\newblock
  Phys. Rev. \textbf{C23}, 645--656 (1981).

\bibitem{Baye.1985.NPA}
  D.~Baye, and P.~Descouvemont,
\newblock
  \emph{Microscopic description of nucleus-nucleus bremsstrahlung},
\newblock
  Nucl. Phys. \textbf{A443}, 302--320 (1985).

\bibitem{Liu.1990.PRC.v41}
  Q.~K.~K.~Liu, Y.~C.~Tang, and H.~Kanada,
\newblock
  \emph{Microscopic calculation of bremsstrahlung emission in $^{3}{\rm He} + \alpha$ collisions},
\newblock
  Phys. Rev. \textbf{C41} (4), 1401--1416 (1990).

\bibitem{Liu.1990.PRC.v42}
  Q.~K.~K.~Liu, Y.~C.~Tang, and H.~Kanada,
\newblock
  \emph{Microscopic study of $p + \alpha$ bremsstrahlung},
\newblock
  Phys. Rev. \textbf{C42} (5), 1895--1898 (1990).

\bibitem{Baye.1991.NPA}
  D.~Baye, C.~Sauwens, P.~Descouvemont, and S.~Keller,
\newblock
  \emph{Accurate treatment of Coulomb contribution in nucleus-nucleus bremsstrahlung},
\newblock
  Nucl. Phys. \textbf{A529}, 467--484 (1991).

\bibitem{Liu.1992.FBS}
  Q.~K.~K.~Liu, Y.~C.~Tang, and H.~Kanada,
\newblock
  \emph{Microscopic study of $\alpha + \alpha$ bremsstrahlung with resonating-group wave functions},
\newblock
  Few-Body Syst. \textbf{12}, 175--189 (1992).

\bibitem{Dohet-Eraly.2011.JPCS}
  J.~Dohet-Eraly, J.-M.~Sparenberg, and D.~Baye,
\newblock
  \emph{Microscopic calculations of elastic scattering between light nuclei based on a realistic nuclear interaction}
\newblock
  J. Phys.: Conf. Ser. \textbf{321}, 012045 (2011).

\bibitem{Dohet-Eraly.2011.PRC}
  J.~Dohet-Eraly, D.~Baye,
\newblock
  \emph{Microscopic cluster model of $\alpha + n$, $\alpha + p$, $\alpha + {\rm He}^{3}$, and $\alpha + \alpha$ elastic scattering from a realistic effective nuclear interaction},
\newblock
  Phys. Rev. \textbf{C 84}, 014604 (2011).

\bibitem{Dohet-Eraly.2013.PhD}
  J.~Dohet-Eraly,
\newblock
  \emph{Microscopic cluster model of elastic scattering and bremsstrahlung of light nuclei},
\newblock
  PhD thesis (Universite Libre De Bruxelles, 2013).

\bibitem{Dohet-Eraly.2013.JPCS}
  J.~Dohet-Eraly, D.~Baye, and P.~Descouvemont,
\newblock
  \emph{Microscopic description of $\alpha + \alpha$ bremsstrahlung from a realistic nucleon-nucleon interaction}
\newblock
  J. Phys.: Conf. Ser. \textbf{436}, 012030 (2013).

\bibitem{Dohet-Eraly.2013.PRC}
  J.~Dohet-Eraly, and D.~Baye,
\newblock
  \emph{Siegert approach within a microscopic description of nucleus-nucleus bremsstrahlung}
\newblock
  Phys. Rev. \textbf{C 88}, 024602 (2013).

\bibitem{Dohet-Eraly.2014.PRC.v89}
   J.~Dohet-Eraly,
\newblock
  \emph{Microscopic description of $\alpha + N$ bremsstrahlung by a Siegert approach},
\newblock
  Phys. Rev. C \textbf{89}, 024617 (2014).

\bibitem{Dohet-Eraly.2014.PRC.v90}
   J.~Dohet-Eraly, and D.~Baye,
\newblock
  \emph{Comparison of potential models of nucleus-nucleus bremsstrahlung},
\newblock
  Phys. Rev. C \textbf{90}, 034611 (2014).
\bibitem{Stoks.1993.PRC}
   V.~G.~J.~Stoks, R.~A.~M.~Klomp, M.~C.~M.~Rentmeester, and J.~J.~de~Swart,
\newblock
  Phys. Rev. C \textbf{48}, 792 (1993).

\bibitem{Maydanyuk.2003.PTP}
  S.~P.~Maydanyuk, V.~S.~Olkhovsky,
\newblock
  \emph{Does sub-barrier bremsstrahlung in $\alpha$ decay of $^{210}\mbox{Po}$ exist?}
\newblock
  Prog. Theor. Phys. \textbf{109} (2), 203--211 (2003);
\newblock
  nucl-th/0404090.

\bibitem{Maydanyuk.2006.EPJA}
  S.~P.~Maydanyuk and V.~S.~Olkhovsky,
\newblock
 \emph{Angular analysis of bremsstrahlung in $\alpha$ decay},
\newblock
  Europ. Phys. Journ. \textbf{A28} (3), 283--294 (2006),
\newblock
  nucl-th/0408022.

\bibitem{Maydanyuk.2008.EPJA}
  G.~Giardina, G.~Fazio, G.~Mandaglio, M.~Manganaro,
  C.~Sacc\'{a}, N.~V.~Eremin, A.~A.~Paskhalov, D.~A.~Smirnov, S.~P.~Maydanyuk, and V.~S.~Olkhovsky,
\newblock
  \emph{Bremsstrahlung emission accompanying  alpha-decay of $^{214}\mbox{\rm Po}$},
\newblock
  Europ. Phys. Journ. \textbf{A36} (1), 31--36 (2008).

\bibitem{Giardina.2008.MPLA}
  G.~Giardina, G.~Fazio, G.~Mandaglio, M.~Manganaro,
  S.~P.~Maydanyuk, V.~S.~Olkhovsky, N.~V.~Eremin, A.~A.~Paskhalov, D.~A.~Smirnov and C.~Sacc\'{a},
\newblock
  \emph{Bremsstrahlung emission during $\alpha$ decay of $^{226}\mbox{\rm Ra}$},
\newblock
  Mod. Phys. Lett. \textbf{A23} (31), 2651--2663 (2008);
\newblock
  arxiv:~0804.2640.

\bibitem{Maydanyuk.2009.NPA}
  S.~P.~Maydanyuk, V.~S.~Olkhovsky, G.~Giardina, G.~Fazio, G.~Mandaglio and M.~Manganaro,
\newblock
  Nucl.~Phys. \textbf{A823}, 3 (2009).

\bibitem{Maydanyuk.2009.JPS}
  S.~P.~Maydanyuk,
\newblock
  \emph{Multipolar approach for description of bremsstrahlung during $\alpha$ decay},
\newblock
  Jour. Phys. Study. \textbf{13} (3), 3201 (2009).

\bibitem{Maydanyuk.2009.TONPPJ}
  S.~P.~Maydanyuk,
\newblock
  \emph{Multipolar approach for description of bremsstrahlung during $\alpha$ decay and unified formula of the bremsstrahlung probability},
\newblock
  Open Nucl. Part. Phys. J. \textbf{2}, 17--33 (2009) [open access].

\bibitem{Maydanyuk.2010.PRC}
  S.~P.~Maydanyuk, V.~S.~Olkhovsky, G.~Mandaglio, M.~Manganaro, G.~Fazio and G.~Giardina,
\newblock
  \emph{Bremsstrahlung emission of high energy accompanying spontaneous of $^{252}{\rm Cf}$},
\newblock
  Phys. Rev. \textbf{C82}, 014602 (2010).

\bibitem{Maydanyuk.2011.JPG}
  S.~P.~Maydanyuk,
\newblock
  Jour. Phys. \textbf{G38} (8), 085106 (2011).

\bibitem{Maydanyuk.2011.JPCS}
  S.~P.~Maydanyuk, V.~S.~Olkhovsky, G.~Mandaglio, M.~Manganaro, G.~Fazio and G.~Giardina,
\newblock
   \emph{Bremsstrahlung emission of photons accompanying ternary fission of $^{252}{\rm Cf}$},
\newblock
  Journ. Phys.: Conf. Ser. \textbf{282}, 012016 (2011).

\bibitem{Maydanyuk.2012.PRC}
  S.~P.~Maydanyuk,
\newblock
  \emph{Model of the bremsstrahlung emission accompanying interactions between protons and nuclei from low up to intermediate energies: role of magnetic emission},
\newblock
  Phys. Rev. \textbf{C86}, 014618 (2012), arXiv:1203.1498.

\bibitem{Maydanyuk_Zhang.2015.PRC}
  S.~P.~Maydanyuk, P.M. Zhang,
\newblock
  \emph{A new approach to determine the proton-nucleus interactions from the experimental bremsstrahlung data},
\newblock
  Phys. Rev. \textbf{C91}, 024605 (2015), arXiv:1309.2784.

\bibitem{Maydanyuk_Zhang_Zou.2016.PRC}
  S.~P.~Maydanyuk, P.M. Zhang, L.-P.~Zou,
\newblock
  \emph{New approach for obtaining information on the many-nucleon structure in $\alpha$ decay from accompanying bremsstrahlung emission},
\newblock
  Phys. Rev. \textbf{C93}, 014617 (2016), arXiv:1505.01029.
\bibitem{Maydanyuk_Zhang.2015.NPA}
  S.~P.~Maydanyuk, P.-M.~Zhang, and S.~V.~Belchikov,
\newblock
  \emph{Quantum design using a multiple internal reflections method in a study of fusion processes in the capture of alpha-particles by nuclei},
\newblock
  Nucl. Phys. A \textbf{940}, 89--118 (2015);
\newblock
  arxiv:1504.00567.
\bibitem{Zakhariev.1985.book}
  B.~N.~Zahariev, A.~A.~Suzko,
\newblock
  \emph{Potentiali i kvantovoye rasseyaniye: Pryamaya i obratnaya zadachi}
\newblock
  (Moskva, Energoatomizdat, 1985), 224~pp.
\bibitem{Boie.2007.PRL}
   H.~Boie, H.~Scheit, U.~D.~Jentschura, F.~K\"{o}ck,
   M.~Lauer, A.~I.~Milstein, I.~S.~Terekhov, and D.~Schwalm,
\newblock
  \emph{Bremsstrahlung in $\alpha$ decay reexamined},
\newblock
  Phys. Rev. Lett. \textbf{99}, 022505 (2007);
\newblock
  arXiv:0706.2109.

\bibitem{Boie.2009.PhD}
  H.~Boie,
\newblock
  \textit{Bremsstrahlung emission probability in the $\alpha$ decay of $^{210}{\rm Po}$},
\newblock
  PhD thesis (Ruperto-Carola University of Heidelberg, Germany, 2009), 193~p.

\bibitem{D'Arrigo.1994.PHLTA}
  A.~D'Arrigo, N.~V.~Eremin, G.~Fazio, G.~Giardina, M.~G.~Glotova, T.~V.~Klochko, M.~Sacchi and A.~Taccone,
\newblock
  \emph{Investigation of bremsstrahlung emission in $\alpha$ decay of heavy nuclei},
\newblock
   Phys. Lett. \textbf{B332}, 25--30 (1994).

\bibitem{Kasagi.1997.JPHGB}
  J.~Kasagi, H.~Yamazaki, N.~Kasajima, T.~Ohtsuki and H.~Yuki,
\newblock
  \emph{Bremsstrahlung emission in $\alpha$ decay and tunneling motion of $\alpha$ particle},
\newblock
  Journ. Phys. \textbf{G 23}, 1451--1457 (1997).

\bibitem{Kasagi.1997.PRLTA}
  J.~Kasagi, H.~Yamazaki, N. Kasajima, T.~Ohtsuki and H.~Yuki,
\newblock
  \emph{Bremsstrahlung in $\alpha$ decay of $^{210}\mbox{Po}$: do $\alpha$ particles emit photons in tunneling?},
\newblock
  Phys. Rev. Lett. \textbf{79}, 371--374 (1997).


\bibitem{Batkin.1986.SJNCA}
  I.~S.~Batkin, I.~V.~Kopytin and T.~A.~Churakova,
\newblock
  Yad. Fiz. (Sov. Journ. Nucl. Phys.) \textbf{44}, 1454--1458 (1986).

\bibitem{Dyakonov.1996.PRLTA}
  M.~I.~Dyakonov, I.~V.~Gornyi,
\newblock
  \emph{Electromagnetic radiation by a tunneling charge},
\newblock
  Phys. Rev. Lett. \textbf{76}, 3542--3545 (1996).

\bibitem{Papenbrock.1998.PRLTA}
  T.~Papenbrock, G.~F.~Bertsch,
\newblock
  \emph{Bremsstrahlung in $\alpha$ decay},
\newblock
  Phys. Rev. Lett. \textbf{80}, 4141--4144 (1998);
\newblock
  nucl-th/9801044.

\bibitem{Tkalya.1999.JETP}
  E.~V.~Tkalya,
\newblock
  Zh. Eksp. Teor. Fiz. \textbf{116}, 390 (1999)
  [Translation: Sov. Phys. JETP \textbf{89} (1999) 208].

\bibitem{Tkalya.1999.PHRVA}
  E.~V.~Tkalya,
\newblock
  \emph{Bremsstrahlung in $\alpha$ decay and ``interference of space regions''},
\newblock
  Phys.~Rev. \textbf{C60}, 054612 (1999).

\bibitem{Bertulani.1999.PHRVA}
  C.~A.~Bertulani, D.~T.~de~Paula and V.~G.~Zelevinsky,
\newblock
  \emph{Bremsstrahlung radiation by a tunneling particle: A time-dependent description},
\newblock
  Phys. Rev. \textbf{C60}, 031602 (1999);
\newblock
  nucl-ex/9812009.

\bibitem{Takigawa.1999.PHRVA}
  N.~Takigawa, Y.~Nozawa, K.~Hagino, A.~Ono and D.~M.~Brink,
\newblock
  \emph{Bremsstrahlung in $\alpha$  decay},
\newblock
  Phys. Rev. \textbf{C59}, R593--R597 (1999);
  nucl-th/9809001.

\bibitem{Flambaum.1999.PRLTA}
  V.~V.~Flambaum and V.~G.~Zelevinsky,
\newblock
  \emph{Quantum M\"{u}nchhausen effect in tunneling},
\newblock
  Phys. Rev. Lett. \textbf{83}, 3108--3111 (1999);
\newblock
  nucl-th/9812076.

\bibitem{Dyakonov.1999.PHRVA}
  M.~I.~Dyakonov,
\newblock
  \emph{Bremsstrahlung spectrum in $\alpha$ decay};
\newblock
  Phys. Rev. \textbf{C60}, 037602 (1999);
  nucl-th/9903016.

\bibitem{So_Kim.2000.JKPS}
  W.~So and Y.~Kim,
\newblock
  \emph{Energy and charge dependency for bremsstrahlung in $\alpha$ decay},
\newblock
  Journ. Korean Phys. Soc. \textbf{37}, 202--208 (2000).

\bibitem{Misicu.2001.JPHGB}
  S.~Misicu, M.~Rizea and W.~Greiner,
\newblock
  \emph{Emission of electromagnetic radiation in $\alpha$ decay},
\newblock
  Journ. Phys. G \textbf{27}, 993--1003 (2001).

\bibitem{Dijk.2003.FBSSE}
  W.~van~Dijk and Y.~Nogami,
\newblock
  \emph{Model study of bremsstrahlung in alpha decay},
\newblock
  Few-body systems Supplement \textbf{14}, 229--232 (2003).

\bibitem{Ohtsuki.2006.CzJP} 
  T.~Ohtsuki, H.~Yuki, K.~Hirose, T.~Mitsugashira,
\newblock
  \emph{Status of the electron accelerator for radioanalytical studies at Tohoku University},
\newblock
  Czech. Journ. Phys. \textbf{56},  D391--D398 (2006).

\bibitem{Amusia.2007.JETP}
  M.~Ya.~Amusia, B.~A.~Zon, and I.~Yu.~Kretinin,
\newblock
  JETP \textbf{105}, 343--346 (2007).

\bibitem{Jentschura.2008.PRC}
  U.~D.~Jentschura, A.~I.~Milstein, I.~S.~Terekhov, H.~Boie, H.~Scheit, and D.~Schwalm,
\newblock
  \emph{Quasiclassical description of bremsstrahlung accompanying $\alpha$  decay including quadrupole radiation},
\newblock
   Phys. Rev. \textbf{C77}, 014611 (2008).
\bibitem{Edington.1966.NP}
  J.~Edington, and B.~Rose,
\newblock
  \emph{Nuclear bremsstrahlung from 140 MeV protonsOriginal},
\newblock
  Nucl. Phys. \textbf{89}, 523 (1966).

\bibitem{Koehler.1967.PRL}
  P.~F.~M.~Koehler, K.~W.~Rothe, and E.~H.~Thorndike,
\newblock
  \emph{Neutron-proton bremsstrahlung at 197~MeV},
\newblock
  Phys. Rev. Lett. \textbf{18}, 933 (1967).

\bibitem{Kwato_Njock.1988.PLB}
  M.~Kwato~Njock, M.~Maurel, H.~Nifenecker, J.~Pinston, F.~Schussler, D.~Barneoud, S.~Drissi,
  J.~Kern, and J.~P.~Vorlet,
\newblock
  \emph{Nuclear bremsstrahlung production in proton-nucleus reactions at 72~MeV},
\newblock
  Phys. Lett. \textbf{B207}, 269 (1988).

\bibitem{Pinston.1989.PLB}
  J.~A.~Pinston, D.~Barneoud, V.~Bellini, S.~Drissi, J.~Guillot, J.~Julien, M.~Kwato~Njock,
  H.~Nifenecker, M.~Maurel, F.~Schussler, and J.~P.~Vorlet,
\newblock
  \emph{Nuclear bremsstrahlung production in proton-nucleus reactions at 168 and 200 MeV},
\newblock
  Phys. Lett. \textbf{B 218}, 128 (1989).

\bibitem{Pinston.1990.PLB}
  J.~A.~Pinston, D.~Barneoud, V.~Bellini, S.~Drissi, J.~Guillot, J.~Julien,
  H.~Nifenecker, and F.~Schussler,
\newblock
  \emph{Proton-deuterium bremsstrahlung at 200~MeV},
\newblock
  Phys. Lett. \textbf{B 249}, 402 (1990).

\bibitem{Clayton.1991.PhD}
  J.~E.~Clayton,
\newblock
  \emph{High energy gamma ray production in proton induced reactions at energies of 104, 145, and 195 MeV},
\newblock
  PhD thesis (Michigan State University, 1991).

\bibitem{Clayton.1992.PRC.p1810}
  J.~Clayton, W.~Benenson, M.~Cronqvist, R.~Fox, D.~Krofcheck, R.~Pfaff,
  T.~Reposeur, J.~D.~Stevenson, J.~S.~Winfield, B.~Young,
  M.~F.~Mohar, C.~Bloch, and D.~E.~Fields,
\newblock
  \emph{Proton-deuteron bremsstrahlung at 145 and 195 MeV},
\newblock
  Phys. Rev. \textbf{C45}, 1810 (1992).

\bibitem{Clayton.1992.PRC.p1815}
  J.~Clayton, W.~Benenson, M.~Cronqvist, R.~Fox, D.~Krofcheck, R.~Pfaff,
  T.~Reposeur, J.~D.~Stevenson, J.~S.~Winfield, B.~Young,
  M.~F.~Mohar, C.~Bloch, and D.~E.~Fields,
\newblock
  \emph{High energy gamma ray production in proton-induced reactions at 104, 145, and 195 MeV},
\newblock
  Phys. Rev. \textbf{C45}, 1815 (1992).

\bibitem{Chakrabarty.1999.PRC}
  D.~R.~Chakrabarty, V.~M.~Datar, Y.~K.~Agarwal, C.~V.~K.~Baba, M.~S.~Samant,
  I.~Mazumdar, A.~K.~Sinha, and P.~Sugathan,
\newblock
  \emph{Hard photon production in $p + ^{197}{\rm Au}$ reaction at $E_{p} =27$~MeV},
\newblock
  Phys. Rev. \textbf{C60}, 024606 (1999).

\bibitem{Goethem.2002.PRL}
  M.~J.~van~Goethem, L.~Aphecetche, J.~C.~S.~Bacelar, H.~Delagrange, J.~Diaz, D.~d'Enterria, M.~Hoefman,
  R.~Holzmann, H.~Huisman, N.~Kalantar-Nayestanaki, A.~Kugler, H.~L\"{o}hner, G.~Martinez,
  J.~G.~Messchendorp, R.~W.~Ostendorf, S.~Schadmand, R.~H.~Siemssen, R.~S.~Simon,
  Y.~Schutz, R.~Turrisi, M.~Volkerts, V.~Wagner, and H.~W.~Wilschut,
\newblock
  \textit{Suppresion of soft nuclear bremsstrahlung in proton-nucleus collisions},
\newblock
  Phys. Rev. Lett. \textbf{88}, 122302 (2002).
\bibitem{Kamanin.1989.PEPAN}
  V.~V.~Kamanin, A.~Kugler, Yu.~E.~Penionzhkevich, I.~S.~Вatkin, I.~V.~Коруtin,
\newblock
  Phys. El. Part. At. Nucl. \textbf{20}, 743--829 (1989).

\bibitem{Pluiko.1987.PEPAN}
  V.~A.~Pluyko, V.~A.~Poyarkov
\newblock
  Phys. El. Part. At. Nucl. \textbf{18}, 374--418 (1987).

\bibitem{Ploeg.1995.PRC}
  H.~van~der Ploeg, J.~C.~S.~Bacelar, A.~Buda, C.~R.~Laurens, and A.~van~der~Woude,
\newblock
  Phys.~Rev. \textbf{C52}, 1915 (1995).

\bibitem{Kasagi.1989.JPSJ}
  J.~Kasagi, H.~Hama, K.~Yoschida et al.,
\newblock
  Journ. Phys. Soc. Jpn. Suppl. \textbf{58}, 620 (1989).

\bibitem{Luke.1991.PRC}
  S.~J.~Luke, C.~A.~Gossett, R.~Vandenbosch,
\newblock
  Phys. Rev. \textbf{C44}, 1548 (1991).

\bibitem{Hofman.1993.PRC}
  D.~J.~Hofman, B.~B.~Back, C.~P.~Montoya, S.~Schadmand, R.~Varma, and P.~Paul,
\newblock
  Phys.~Rev. \textbf{C47}, 1103 (1993).

\bibitem{Varlachev.2007.BRASP}
  V.~A.~Varlachev, G.~N.~Dudkin, V.~N.~Padalko,
\newblock
  Bull. Russ. Acad. Sci.: Phys. \textbf{71}, 1635--1639 (2007).

\bibitem{Eremin.2010.IJMPE}
  N.~V.~Eremin, A.~A.~Paskhalov, S.~S.~Markochev, E.~A.~Tsvetkov, G.~Mandaglio, M.~Manganaro,
  G.~Fazio, G.~Giardina and M.~V.~Romaniuk,
\newblock
  \emph{New experimental method of investigation the rare nuclear transformations accompanyimg atomic processes: Bremsstrahlung emission in spontaneous fission of $^{252}{\rm Cf}$},
\newblock
  Int. J. Mod. Phys. \textbf{E19}, 1183 (2010).

\bibitem{Pandit.2010.PLB}
  Deepak~Pandit, S.~Mukhopadhyay, Srijit~Bhattacharya, Surajit~Pal, A.~De and S.~R.~Banerjee,
\newblock
  \emph{Coherent bremsstrahlung and GDR width from $^{252}{\rm Cf}$ cold fission},
\newblock
  Phys. Lett. \textbf{B690} (5), 473--476 (2010).
\bibitem{Kurgalin.2001.IRAN}
  S.~D.~Kurgalin, Yu.~M.~Chuvilskiy, and T.~A.~Churakova,
\newblock
  Izv. Acad. Nauk: Ser. Fiz. \textbf{65}, 672 (2001) [in Russian].
\bibitem{Wolfli.1971.PRL}
  W.~W\"{o}lfli, J.~Hall, and R.~M\"{u}ller,
\newblock
  \emph{Bremsstrahlung in proton -- $\alpha$-particle scattering},
\newblock
  Phys. Rev. Lett. \textbf{27}, 271 (1971).

\bibitem{Peyer.1971.PLB}
  U.~Peyer, J.~Hall, R.~M\"{u}ller, M.~Sutter, and W.~W\"{o}lfli,
\newblock
  \emph{Bremsstrahlung in $\alpha +\alpha$ scattering},
\newblock
  Phys. Lett. \textbf{B41}, 151 (1972).

\bibitem{Frois.1973.PRC}
  B.~Frois, J.~Birchall, C.~R.~Lamontagne, U.~von~Moellendorff, R.~Roy, R.~J.~Slobodrian,
\newblock
  \emph{Bremsstrahlung in the $\alpha - \alpha$ and ${\rm He}^{3} - \alpha$ interactions},
\newblock
  Phys. Rev. \textbf{C 8}, 2132 (1973).

\bibitem{Anzelon.1975.NPA}
  G.~A.~Anzelon, I.~Slaus, S.~Y.~Tin, W.~T.~H.~Van~Oers,
  R.~M.~Eisberg, M.~Makino, C.~N.~Waddell, and M.~B.~Epstein,
\newblock
  \emph{Bremsstrahlung in proton - $\alpha$ scattering},
\newblock
  Nucl. Phys. \textbf{A255}, 250-266 (1975).

\bibitem{Wildermuth.1977.book}
  K.~Wildermuth and Y.~C.~Tang,
\newblock
  \emph{A Unifierd Theory of the Nucleus},
\newblock
  (Vieweg, 1977).

\bibitem{Tang.1978.PR}
  Y.~C.~Tang, M.~LeMere, and D.~R.~Thompson,
\newblock
  \emph{Resonating-group method for nuclear many-body problems},
\newblock
  Physics Reports~\textbf{47}, 167--223 (1978).

\bibitem{Tang.1981.lectures}
  Y.~C.~Tang,
\newblock
  in \emph{Topics in Nuclear Physics II},
\newblock
  Lecture Notes in Physics,
  Vol.~\textbf{145} (Springer, Berlin, 1981), pp.~571--692.



\bibitem{Horiuchi.1977.PTPS}
  H.~Horiuchi,
\newblock
  Prog. Theor. Phys. Suppl. \textbf{62}, 90 (1977).


\bibitem{Eberhard.1979.PRL}
  K.~A.~Eberhard, Ch.~Appel, R.~Bangert, L.~Cleemann, J.~Eberth, and V.~Zobel,
\newblock
  \emph{Fusion cross sections for $\alpha + ^{40, 44}{\rm Ca}$ and the problem of anomalous large-angle scattering},
\newblock
  Phys. Rev. Lett. \textbf{43}, 107--110 (1979).
\bibitem{Arndt.1971.PRC}
  R.~A.~Arndt, L.~D.~Roper, and R.~L.~Shotwell,
\newblock
  Phys. Rev. \textbf{C3}, 2100 (1971).


\bibitem{Kopitin.1997.YF}
  I.~V.~Kopitin, M.~A.~Dolgopolov, T.~A.~Churakova, A.~S.~Kornev,
\newblock
  Phys. At. Nucl. \textbf{60} (5), 776--785 (1997)
  [Rus. ed.:  Yad. Fiz. \textbf{60} (5), 869--879 (1997)].

\bibitem{Steshenko.1971.YF}
  A.~I.~Steshenko, G.~F.~Filipov
\newblock
  Yad. Fiz. \textbf{14}, 715 (1971)
  [Journ. Nucl. Phys. \textbf{14}, 715 (1971)].

\bibitem{Becchetti.1969.PR}
  F.~D.~Becchetti, Jr., and G.~W.~Greenless,
\newblock
  \emph{Nucleon-nucleon optical-model parameters, $A > 40$, $E < 50$ MeV},
\newblock
  Phys. Rev. \textbf{182} (4), 1190--1209 (1969).

\bibitem{Reichstein.1970.NPA}
  I.~Reichstein, and Y.~C.~Tang,
\newblock
  Nucl. Phys. \textbf{A158}, 529 (1970).
\bibitem{Bertholet.1987.NPA}
  R.~Bertholet, M.~Kwato Njock, M.~Maurel, E.~Monnand, H.~Nifenecker, P.~Perrin,
  J.~A.~Pinston, F.~Schussler D.~Barneoud, C.~Guet, and Y.~Schutz,
\newblock
  \emph{High energy gamma-ray production from 44 MeV/A 86Kr bombardment on nuclei},
\newblock
  Nucl. Phys. \textbf{A474}, 541-556 (1987).
\bibitem{Varlachev.2005.JETPL}
  V.~A.~Varlachev, G.~N.~Dudkin, and V.~N.~Padalko,
\newblock
  \emph{Does the coherent bremsstrahlung of fission fragments exist?},
\newblock
  Journ. Exp. Theor. Phys. Lett. \textbf{82} (7), 390--393 (2005).

\end{thebibliography}
\end{document}